\documentclass[10pt,letterpaper]{article}
\usepackage[margin=1in]{geometry}

\usepackage{multirow}
\usepackage{epsfig}
\usepackage{paralist}
\usepackage[center]{subfigure}
\usepackage[font=normalsize]{caption}
\usepackage{enumitem}
\usepackage{color}
\usepackage{xspace}
\usepackage{amsmath}
\usepackage{bbold}
\usepackage{mathtools}
\usepackage{comment}
\usepackage{microtype}
\usepackage{booktabs}
\usepackage[misc]{ifsym}
\usepackage{amssymb}
\usepackage[hidelinks]{hyperref}
\usepackage{algorithmic}
\usepackage[ruled,vlined,linesnumbered,resetcount,noend]{algorithm2e}

\usepackage{xcolor}
\include{pythonlisting}
\SetKwInput{KwInput}{Input}                % Set the Input
\SetKwInput{KwOutput}{Output}              % set the Output
\SetKwComment{Comment}{$\triangleright$\ }{}

\setlist[itemize]{leftmargin=*}

\newtheorem{theorem}{Theorem}
\newtheorem{proof}{Proof}

\newcommand{\SummaryGraph}{G'}
\newcommand{\WeightedSummaryGraph}{\overline{G}}
\newcommand{\UnweightedSummaryGraph}{\tilde{G}}
\newcommand{\WeightFunction}{\omega}
\newcommand{\SupernodeI}{S_i}
\newcommand{\SupernodeJ}{S_j}
\newcommand{\PIAB}{\Pi_{AB}}
\newcommand{\EAB}{E_{AB}}
\newcommand{\PIIJ}{\Pi_{S_{i}S_{j}}}
\newcommand{\WeightFunctionAB}{\omega_{AB}}
\newcommand{\WeightFunctionIJ}{\omega_{S_iS_j}}
\newcommand{\ReconstructedWeightFunctionIJ}{\hat{\omega}_{ij}}
\newcommand{\Wmax}{\omega_{max}}
\newcommand{\ReconstructedAdjacencyMatrix}{A'}
\newcommand{\WeightedReconstructedGraph}{\hat{G}}
\newcommand{\WeightedReconstructedAdjacencyMatrix}{\hat{A}}
\newcommand{\WeightFunctionReconstructed}{\hat{\omega}}
\newcommand{\UnweightedReconstructedGraph}{\check{G}}
\newcommand{\UnweightedReconstructedAdjacencyMatrix}{\check{A}}

\newcommand{\kGrass}{\textsc{k-Grass}\xspace}
\newcommand{\kGrassUnweighted}{\textsc{k-Grass (Unweighted)}\xspace}
\newcommand{\SSumM}{\textsc{SSumM}\xspace}
\newcommand{\SSumMUnweighted}{\textsc{SSumM (Unweighted)}\xspace}
\newcommand{\MoSSoLossy}{\textsc{MoSSo-Lossy}\xspace}

\newcommand{\MoSSoLossyUnweighted}{\textsc{MoSSo-Lossy (Unweighted)}\xspace}
\newcommand{\MoSSo}{\textsc{MoSSo}\xspace}

\newcommand{\smallsection}[1]{{\vspace{0.02in} \noindent {\bf{\underline{\smash{#1}}}}}}
\usepackage{graphicx}
\begin{document}

\title{Are Edge Weights in Summary Graphs Useful? - A Comparative Study}
%
%\titlerunning{Are Edge Weights in Summary Graphs Useful? - A Comparative Study}

\author{Shinhwan Kang,\textsuperscript{1}
Kyuhan Lee,\textsuperscript{1} and
Kijung Shin\textsuperscript{1,2}
\date{\textsuperscript{1}Kim Jaechul Graduate School of AI and \textsuperscript{2}School of Electrical Engineering \\ KAIST \\ Seoul, South Korea\\
\{shinhwan.kang, kyuhan.lee, kijungs\}@kaist.ac.kr}
}
%
%\authorrunning{Kang et al.}

\maketitle              % typeset the header of the contribution

\begin{abstract}
    \textit{Which one is better between two representative graph summarization models with and without edge weights?}
From web graphs to online social networks, large graphs are everywhere. Graph summarization, which is an effective graph compression technique, aims to find a compact \textit{summary graph} that accurately represents a given large graph.
Two versions of the problem, where one allows edge weights in summary graphs and the other does not, have been studied in parallel without direct comparison between their underlying representation models.
In this work, we conduct a systematic comparison by extending three search algorithms to both models and evaluating their outputs on eight datasets in five aspects: (a) reconstruction error, (b) error in node importance, (c) error in node proximity, (d) the size of reconstructed graphs, and (e) compression ratios.
Surprisingly, using unweighted summary graphs leads to outputs significantly better in all the aspects
than using weighted ones, and this finding is supported theoretically.
Notably, we show that
a state-of-the-art algorithm can be improved substantially
(specifically,
$8.2\times$, $7.8\times$, and $5.9\times$ in terms of (a), (b), and (c), respectively, when (e) is fixed) based on the observation.    
\end{abstract}

\begin{figure*}[t]
    \centering
    \includegraphics[width=1\textwidth]{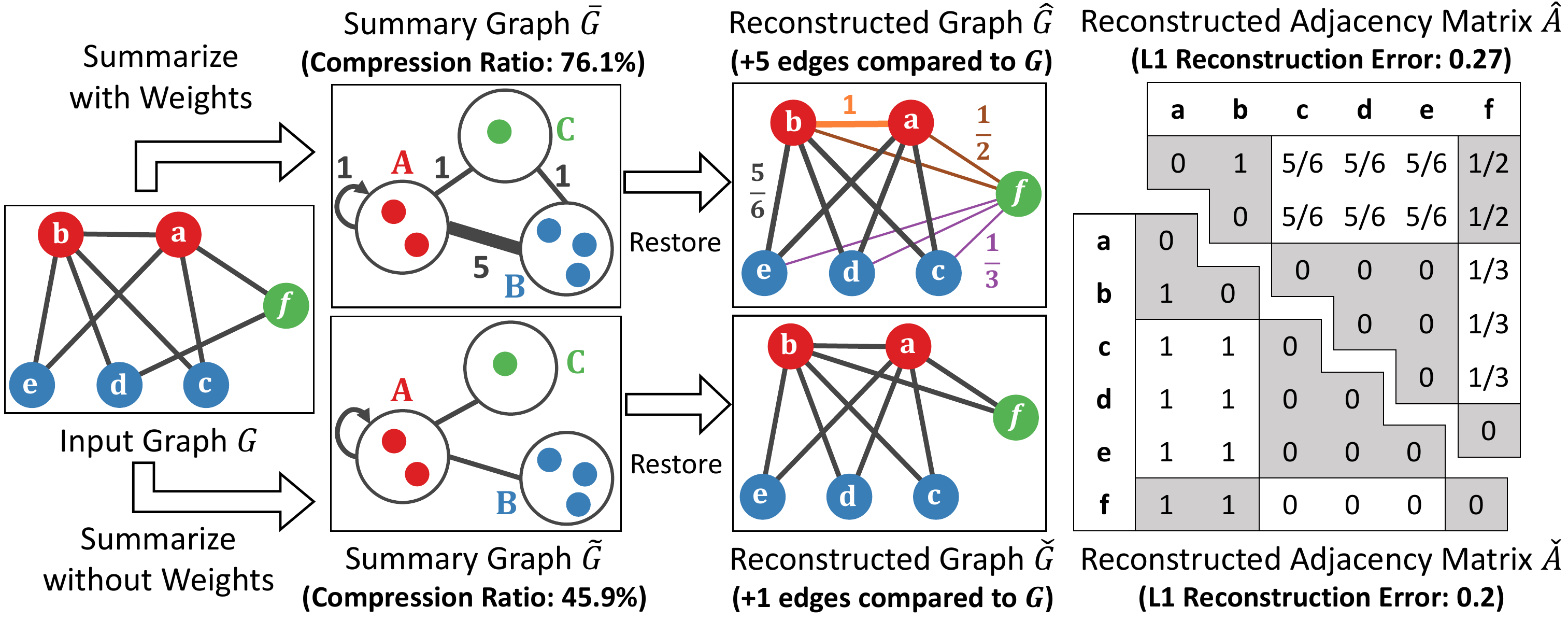} 
    \caption{Examples of weighted and unweighted graph summarization and reconstruction. The summary graph $\UnweightedSummaryGraph$ without edge weights is more concise with smaller reconstruction error and reconstruction size than $\WeightedSummaryGraph$ with edge weights.
    \label{fig:SummarizationandReconstruction}
    }
\end{figure*}

\section{Introduction and Related Works}
\label{sec:intro}

%\red{PageRank Citation?, Four/Five aespect?}

Relationships between objects, such as friendships in online social networks, co-appearance of tags, and hyperlinks between web pages, are universal. They are naturally represented as graphs, whose sizes have grown at a tremendous rate due to advances in web technology. For example, the number of web pages (i.e., nodes in web graphs) increased by about $500\times$ in the past two decades. % over the last two decades due to the universalization of the Web.

Graph compression is a useful technique for efficient  utilization of such large graphs. % such as web graphs by representing them in a compact representation.
Many techniques have been developed for various purposes, including storage \cite{boldi2004webgraph,chierichetti2009compressing,dhulipala2016compressing,ko2020incremental,lee2020ssumm,lim2014slashburn,navlakha2008graph,rossi2018graphzip,shin2019sweg},
% \red{Many techniques have been developed for various purposes, including storage \cite{ko2020incremental,lee2020ssumm,navlakha2008graph,shin2019sweg},}
query processing \cite{buehrer2008scalable,kang2022personalized,lefevre2010grass,riondato2017graph},
% \red{query processing \cite{lefevre2010grass,riondato2017graph},}
influence analysis \cite{mathioudakis2011sparsification,mehmood2013csi},
pattern mining \cite{koutra2014vog,shah2015timecrunch},
% anomaly/outlier detection \cite{belth2020normal,davis2011detecting,noble2003graph}, 
anomaly/outlier detection \cite{belth2020normal,davis2011detecting}, 
%community detection \cite{blondel2008fast},
privacy preservation \cite{shoaran2013zero,sui2018privacy}, 
visualization \cite{dunne2013motif,dwyer2013edge}, and 
% representation learning \cite{chen2018harp,fahrbach2020faster,zhou2021dpgs}.
representation learning \cite{fahrbach2020faster,zhou2021dpgs}.
% compressing various types of graphs, including labeled graphs \cite{fan2012query,feder1995clique,}, weighted graphs \cite{henecka2015lossy,liu2012compressing,toivonen2011compression}, dynamic graphs \cite{dolgorsuren2019starzip,liang2020reachability,mustafa2019dynamic,tsalouchidou2018scalable}, and
% domain-specific graphs \cite{boldi2008large,buehrer2008scalable,chierichetti2009compressing,collins2017single,pradhan2007second}. 
We refer to  surveys
\cite{besta2018survey,liu2018graph} for details of them. Their common goal is to find a compact representation that exactly or approximately describes a given graph. 

Among them, we focus on \textit{graph summarization} \cite{beg2018scalable,kang2022personalized,khan2015set,ko2020incremental,lee2020ssumm,lefevre2010grass,navlakha2008graph,riondato2017graph,shin2019sweg}.
% \red{Among them, we focus on \textit{graph summarization} \cite{ko2020incremental,lee2020ssumm,lefevre2010grass,navlakha2008graph,riondato2017graph,shin2019sweg}.}
whose objective is to find a concise \textit{summary graph} $\SummaryGraph$ that accurately describes a given large graph $G$, or equivalently, concise $\SummaryGraph$  from which we can restore a graph close to $G$.\footnote{While we use the term ``graph summarization'' to refer to this specific way of compression, the term has also been used more generally, as surveyed in \cite{liu2018graph}.} 
Each node in $\SummaryGraph$ is interpreted as a group of nodes in $G$, and each edge in $\SummaryGraph$ is interpreted as the presence of edges between all pairs of nodes in two groups. 
Since the output $\SummaryGraph$ is in the form of a graph, other graph compression methods can be applied to $\SummaryGraph$ for further compression \cite{shin2019sweg}. That is, graph summarization can be used as a preprocessing step of other compression methods.
Moreover, a wide range of graph algorithms can be approximately executed on $\SummaryGraph$ without full reconstruction
(see Appendix~\ref{appendix:query} and \cite{riondato2017graph}).

%\red{[TODO: mention query??]}

%connect the two nodes of $\SummaryGraph$. The edge of $\SummaryGraph$ indicates the presence of edges in G in any pair in the two sets (i.e., a set of nodes of $G$). 
%We aim to 
%From $G'$, we can reconstruct a graph 
%We can \textbf{reconstruct} a graph that is the same or some errors as an original graph $G$ using nodes and edges of the obtained summary graph $\SummaryGraph$.

% Graph summarization models can be categorized into two: 
There are two representative graph summarization models: a summary graph with edge weights \cite{beg2018scalable,lee2020ssumm,lefevre2010grass,riondato2017graph} 
% \red{There are two representative graph summarization models: a summary graph with edge weights \cite{lee2020ssumm,lefevre2010grass,riondato2017graph} }
 and one without edge weights  \cite{kang2022personalized,khan2015set,ko2020incremental,navlakha2008graph,shin2019sweg}.
%While graph summarization models can be categorized by various criteria,  Representative one of many standards is whether it is a lossless summarization  or a lossy summarization . Also, whether to use weights or not is an example of many criteria, it is our main issue. 
While the latter is typically used with edge corrections for lossless compression, this work focuses on $G'$.
While a number of search algorithms aiming at finding a high-quality summary graph under a given constraint have been developed for each model, %to the best of our knowledge, 
there was no systematic comparison between the two models.

%experiments between these methods of optimization (i.e., algorithms) exist, but between models have never been conducted.
Which one is better between the two graph summarization models? Are edge weights in summary graphs useful?
%To answer these questions, we conduct 
For a systematic comparison between the two models, we extend three search algorithms \cite{ko2020incremental,lee2020ssumm,lefevre2010grass} to both models and evaluate their outputs in eight real-world graphs in five aspects: (a) reconstruction error, (b) error in node importance \cite{page1999pagerank}, 
(c) error in node proximity \cite{tong2008random},  (d) the number of edges in  reconstructed graphs, and (e) compression ratios.

Counterintuitively, we find out that using unweighted summary graphs gives a significantly better trade-off among (a)-(e) than using weighted ones, regardless of search algorithms and datasets (See Fig.~\ref{fig:SummarizationandReconstruction} for an example).
%Noteworthy, adapting a state-of-the-art algorithm for the weighted model \cite{lee2020ssumm} to the unweighted model leads to $6.3\times$, $7.8\times$, and $2.2\times$ improvements in terms of (a), (b), and (c), when (d) is fixed.
Notably, adapting a state-of-the-art algorithm for the weighted model \cite{lee2020ssumm} to the unweighted model leads to $8.2\times$, $7.8\times$, and $5.9\times$ improvements in terms of (a), (b), and (c) (when (e) is fixed) and $2.2\times$ improvements in terms of (d) (when (a) is similar).

Our contributions are three-fold:
%Comparing two graphs reconstructed from two summary graphs with the same storing cost using bits as a unit, it was shown that weighted models have up to \textbf{6.3$\times$} higher reconstruction errors and up to \textbf{1.5$\times$} PageRank errors compared to unweighted models, and \textbf{2.2$\times$} more edges in reconstructed grpahs in a comparsion of SOTA model. These results show unweighted models have better performance. 
% When having the same reconstruction error, the number of edges reconstructed from the summary graph is lower, we can say that the quality of the summary graph is good.
\begin{itemize}[leftmargin=*]
    \item \textbf{Systematic Comparison:} We conduct a systematic comparison between two extensively-studied graph summarization models using three search algorithms, eight datasets, and five evaluation metrics.
    \item \textbf{Unexpected Observation:} Our comparison leads to a surprising observation that using unweighted models is significantly better than using weighted ones in all considered aspects.
    We support this finding theoretically (see Theorem~\ref{theorem:l1}).
    \item \textbf{Improvement of the State of the Art:} By exploiting the observation, we can improve a state-of-the-art algorithm \cite{lee2020ssumm} substantially
    in all considered aspects (see Figs.~\ref{fig:ExperimentsResult_L1L2}-\ref{fig:ExperimentsResult_NUMRE}). % by exploiting the observation.
    %for the weighted model by adapting it to the unweighted model.
    %\item \textbf{Algorithms Variation for Fair Comparison:} For fair experiments, we extend the existing three algorithms to two models.
    %\item \textbf{Extensive Experiment:} We conduct comparative experiments on data across different sizes and fields.
\end{itemize}
\smallsection{Reproducibility:} 
The source code and the datasets are available at \cite{appendix}.

%The rest of this paper is organized as follows. 
\smallsection{Roadmap:} 
In Sect.~\ref{sec:model}, we introduce graph summarization models. 
In Sect.~\ref{sec:problem}, we define problems and present algorithms. 
In Sect.~\ref{sec:experiments}, we provide empirical results. 
In Sect.~\ref{sec:analysis}, we present theoretical results.
In Sect.~\ref{sec:conclusions}, we offer conclusions.

\section{Graph Summarization Models}
\label{sec:model}

%\begin{table}[]
%	\small
%	\begin{center}
%		\caption{Symbols and Definitions.}
%		\scalebox{0.9}{
%		    \label{tab:symbol}
%			\begin{tabular}{l|l}
%				\toprule 
%				\textbf{Symbol}  & \textbf{Definition}\\
%				\midrule
%				$G = (V,E)$ & input graph with subnodes $V$ and subedges $E$\\
%				$A$ & adjacency matrix of $G$\\
%				\midrule
%				$S_{i}$ & supernode containing the subnode $i$ \\
%				$\PIAB$ & set of possible subedges between supernodes $A$ and $B$\\
%				$\EAB$ & set of subedges between supernodes $A$ and $B$\\
%				% $\PIS$ & set of possible unordered pairs of supernodes\\
%				\midrule
%				$\WeightedSummaryGraph = (S,P, \WeightFunction)$ & weighted summary graph with supernodes $S$, superedges $P$, and weight function $\WeightFunction$\\
%				$\WeightedReconstructedGraph = (V,\check{E})$ &  reconstructed graph with subnodes $V$, subedges $\hat{E}$\\
%				$\WeightedReconstructedAdjacencyMatrix$  &  reconstructed adjacency matrix of $\WeightedReconstructedGraph$\\		
%				\midrule
%				$\UnweightedSummaryGraph = (S,P)$ & unweighted summary graph with supernodes $S$, superedges $P$\\
%				$\UnweightedReconstructedGraph = (V,\hat{E})$ &  reconstructed graph with subnodes $V$, subedges $\hat{E}$\\
%				$\UnweightedReconstructedAdjacencyMatrix$  &  reconstructed adjacency matrix of $\UnweightedReconstructedGraph$\\	
%				\bottomrule
%			\end{tabular}
%		}
%	\end{center}
%\end{table}

\begin{table}[t]
	\begin{center}
		\caption{Symbols and definitions.}
		\scalebox{0.90}{
		    \label{tab:symbol}
			\begin{tabular}{l|l|l|l}
				\toprule 
				\textbf{Symbol}  \ & \ \textbf{Definition} \ & \ \textbf{Symbol}  \ & \ \textbf{Definition}\\
				\midrule
				$V$ \ & \ set of subnodes \ & \ $G = (V,E)$ \ & \ input graph \ \\
				$E$ \ & \ set of subedges \ &  \ $\WeightedSummaryGraph = (S,P, \WeightFunction)$ \ & \ weighted summary graph \\
				$S$ \ & \ set of supernodes \ & \ $\UnweightedSummaryGraph = (S,P)$ \ & \ unweighted summary graph\\
				$P$ & \ set of superedges \ & \ $\WeightedReconstructedGraph = (V,\hat{E}, \WeightFunctionReconstructed)$ \ & \  graph reconstructed from $\WeightedSummaryGraph$ \\
				$\omega$ \ & \ superedge weight function \ & \ $\UnweightedReconstructedGraph = (V,\check{E})$ \ & \  graph reconstructed from $\UnweightedSummaryGraph$ \\
				$\hat{E}, \check{E}$ \ & \ sets of reconstructed edges \  & \ $A$, $\WeightedReconstructedAdjacencyMatrix$, $\UnweightedReconstructedAdjacencyMatrix$  \ & \  adjacency matrix of $G$, $\WeightedReconstructedGraph$, and $\UnweightedReconstructedGraph$ \\
				$\hat{\omega}$ \ & \ subedge weight function \ & \ $\EAB$ \ & \ \# of  subedges between $A,B\in S$\\
				$S_{i}$ \ & \ supernode containing $i\in V$ \ & \ 	$\PIAB$ \ & \ \# of subnode pairs between $A,B\in S$ \\
				\bottomrule
			\end{tabular}
		}
	\end{center}
\end{table}

We introduce weighted and unweighted graph summarization models, which are compared throughout this work.
See Table~\ref{tab:symbol} for frequently-used symbols.

\smallsection{Input Graph.}
Consider an undirected graph $G = (V, E)$ with a set of \textit{subnodes} $V=\{1,\cdots,|V|\}$ and a set of \textit{subedges} $E\subseteq{ V \choose 2 }$. 
We use $A\in \mathbb{R}^{|V|\times|V|}$ to denote its \textit{adjacency matrix}. Each entry $A_{ij}=1$ if $\{i,j\}\in E$ and $A_{ij}=0$ otherwise.

\subsection{Weighted Graph Summarization Model}
\label{sec:model:weighted}

\smallsection{Definition.}
A \textit{weighted summary graph} $\WeightedSummaryGraph = (S, P, \WeightFunction)$ of $G = (V,E)$ consists of a set of \textit{supernodes} $S$, a set  of \textit{superedges} $P$, and a \textit{superedge weight function} \textit{$\WeightFunction$}.
The set $S$ is a partition of $V$. That is, supernodes are disjoint sets of subnodes whose union is $V$.
Each superedge $\{A,B\}\in P$ joins two supernodes $A\in S$ and $B\in S$.
%indicates that there exists a subedge joining subnodes between $A\in S$ and $B\in S$, i.e, there exists $i\in A$ and $j\in B$ where $\{i,j\}\in E$.
The function $\WeightFunction$ takes each superedge $\{A,B\}\in P$ and returns its \textit{weight} $\WeightFunctionAB$, which is equal to $\EAB:=|\{\{i,j\}\in E: i\in A, j\in B\}|$, i.e., the number of subedges in $G$ that join subnodes between $A\in S$ and $B\in S$.

\smallsection{Reconstruction.}
The \textit{reconstructed graph} $\WeightedReconstructedGraph = (V, \hat{E}, \WeightFunctionReconstructed)$ obtained from $\WeightedSummaryGraph = (S, P, \WeightFunction)$ consists of the set of subnodes $V$, the set of reconstructed subedges $\hat{E}\subseteq{ V \choose 2 }$, and a subedge weight function $\WeightFunctionReconstructed$.
%and we denote its adjacency matrix by $\ReconstructedAdjacencyMatrix\in \mathbb{R}^{|V|\times|V|}$. 
If we let $\SupernodeI\in S$ be the supernode containing each subnode $i\in V$ and let $\PIAB$ be the number of possible pairs of subnodes between supernodes $A$ and $B$. That is, $\PIAB:= {|A| \choose 2}$ if $A=B$ and $\PIAB:= |A|\cdot|B|$ otherwise.
The adjacency matrix $\WeightedReconstructedAdjacencyMatrix\in \mathbb{R}^{|V|\times|V|}$ of $\WeightedReconstructedGraph$ is defined as
\begin{equation}
\hat{A}_{ij} = \ReconstructedWeightFunctionIJ := 
    \begin{cases}
        \frac{\WeightFunctionIJ}{\PIIJ}, & \text{if } i\neq j \text{ and }  \{\SupernodeI,\SupernodeJ\}\in P, \\
        0, & \text{otherwise.}
    \end{cases}
    \label{eq:WeightedAdjValue}
\end{equation}
% \red{Note that various graph algorithms can be executed on $\WeightedReconstructedGraph$ without full reconstruction (see Appendix~\ref{appendix:query} and \cite{riondato2017graph}), while we describe the reconstruction of the entire graph $\WeightedReconstructedGraph$ for ease of explanation.}

\subsection{Unweighted Graph Summarization Model}
\label{sec:model:unweighted}

\smallsection{Definition.}
An \textit{unweighted summary graph} $\UnweightedSummaryGraph = (S, P)$ of $G$ consists of a set of supernodes $S$ and a set of superedges $P$. 
Note that, unlike $\WeightedSummaryGraph$, $\UnweightedSummaryGraph$ does not have the superedge weight function $\omega$.

%The only difference with the $\WeightedSummaryGraph$ is that $\WeightFunction$ does not exist. The unweighted summary graph  has a set $S$ of supernodes, a set $P$ of superedges connects two supernodes. The size in bits of $\UnweightedSummaryGraph$ is defined as $2|P|\log_{2}|S| + |V|\log_{2}|S|$. It is the same as the size of $\WeightedSummaryGraph$ minus bits for superedge weights.

\smallsection{Reconstruction.}
The adjacency matrix $\UnweightedReconstructedAdjacencyMatrix\in \mathbb{R}^{|V|\times|V|}$ of 
the graph $\UnweightedReconstructedGraph = (V, \check{E})$ reconstructed from $\UnweightedSummaryGraph$ is defined as
\begin{equation}
\check{A}_{ij} := 
    \begin{cases}
        1, & \text{if } i\neq j \text{ and }  \{\SupernodeI,\SupernodeJ\}\in P \\
        0, & \text{otherwise.}
    \end{cases}
    \label{eq:unweightedAdjValue}
\end{equation}
% While $\UnweightedReconstructedGraph$ is typically used with edge corrections for lossless compression \cite{khan2015set,ko2020incremental,navlakha2008graph,shin2019sweg}, 
% this work focuses on $\UnweightedReconstructedGraph$.
While $\UnweightedReconstructedGraph$ is typically used with edge corrections for lossless compression \cite{ko2020incremental,navlakha2008graph,shin2019sweg}, 
this work focuses on $\UnweightedReconstructedGraph$.

\section{Problem Formulation and Algorithms}
\label{sec:problem}

Based on the graph summarization models, we formulate graph summarization as optimization problems.
Then, we present six search algorithms for the problems.

% \red{We outline the skeleton codes for the six algorithms explained below in Alg~\ref{alg:static}-\ref{alg:dynamic}, and provide details about the codes colored according to algorithms in Table~\ref{tab:algorithmDetail}.}
% The source code of our implementation of all the above algorithms are available at \cite{appendix}.

\subsection{Optimization Problem Formulation}
%\smallsection{Input.}
%Both models consider an undirected graph\ $G = (V,E)$ as an input.

%\smallsection{Problem Definition.}
Given a  graph $G$,  we aim to minimize the difference between a reconstructed adjacency matrix $\ReconstructedAdjacencyMatrix$ (i.e., $\UnweightedReconstructedAdjacencyMatrix$ or $\WeightedReconstructedAdjacencyMatrix$) and the adjacency matrix $A$ of $G$. Specifically, we aim to minimize the $L_p$ reconstruction error, i.e,.
\begin{equation}
\label{eq:ReconstructionError}
RE_{p}(A, \ReconstructedAdjacencyMatrix) := ||A-A'||_{p},
\end{equation}
\noindent while constraining the size of the output summary graph $G'$ (i.e., $\WeightedSummaryGraph$ or $\UnweightedSummaryGraph$) to be at most a given constant. The size can be 
(a) the number of supernodes in $G'$\cite{beg2018scalable,lefevre2010grass,riondato2017graph}, 
% \red{(a) the number of supernodes in $G'$\cite{lefevre2010grass,riondato2017graph},}
(b) the number of superedges in $G'$, or (c) the size of $G'$ in bits \cite{lee2020ssumm}.

\smallsection{Size in Bits of Summary Graphs.}
\label{sec:sizeofsummarygraph}
The size of a weighted summary graph $\WeightedSummaryGraph= (S, P, \WeightFunction)$ in bits is defined as
% \red{The size of $\WeightedSummaryGraph= (S, P, \WeightFunction)$ is defined as}
\begin{equation}
size_{bits}(\WeightedSummaryGraph):= 2|P|\log_{2}|S| + |P|{\log_{2}\Wmax} + |V|\log_{2}|S|,  
\label{eq:size:weighted}
\end{equation}
where $\Wmax$ is the largest superedge weight in $\WeightedSummaryGraph$, and in our experiments in Sect.~\ref{sec:experiments}, $\Wmax<< |S|$.
The three terms on the right side in Eq.~\eqref{eq:size:weighted} correspond to $|P|$ superedges in bits, $|P|$ superedge weights in bits, and the supernode membership of $|V|$ subnodes in bits, respectively.
Similarly, the size of an unweighted summary graph $\UnweightedSummaryGraph$  in bits is defined as
% \red{Similarly, the size of $\UnweightedSummaryGraph$ is defined as}
\begin{equation}
size_{bits}(\UnweightedSummaryGraph):= 2|P|\log_{2}|S|+ |V|\log_{2}|S|. 
\label{eq:size:unweighted}
\end{equation}

%delete
\begin{table}[t!]
	\begin{center}
		\caption{The outlines of the considered search algorithms are given in Algorithms~\ref{alg:static} and \ref{alg:dynamic}, and the details of each algorithm are provided in the table below.  
% 		\red{\textbf{Our implementations of all the search algorithms are available at \cite{appendix}.}}
		\label{tab:Algorithms}\label{tab:algorithmDetail}}
		\small
% 		\begin{tabular}{l|c|c|c|c|c|c}
% 			\toprule 
% 			Algorithm & Outline & \ $\SummaryGraph$ \ & $size()$ & $candidates()$ & $loss()$ & $sparsify?()$  \\
% 			\midrule
% 			k-Grass (W) & Alg.~\ref{alg:static} & $\WeightedSummaryGraph$ &  $|S|$ & $\{S\}$ & Eq.~\eqref{eq:ReconstructionError} & False \\
% 			k-Grass (U) &  Alg.~\ref{alg:static} &$\UnweightedSummaryGraph$ & $|S|$ & $\{S\}$ & Eq.~\eqref{eq:ReconstructionError} & False \\
% 			SSumM (W)& Alg.~\ref{alg:static} & $\WeightedSummaryGraph$ & Eq.~\eqref{eq:size:weighted} & Clusters \cite{buehrer2008scalable} & Eq.~\eqref{eq:SSumMObjective} & True \\
% 			SSumM (U) & Alg.~\ref{alg:static} & $\UnweightedSummaryGraph$ &  Eq.~\eqref{eq:size:unweighted} & Clusters \cite{buehrer2008scalable} & Eq.~\eqref{eq:SSumMUnweightedObjective} & True \\
% % 			MoSSo-Lossy (W)& Alg.~\ref{alg:dynamic} & $\UnweightedSummaryGraph$ & N/A & Clusters \cite{buehrer2008scalable} & Eq.~\eqref{eq:SSumMObjective} & N/A\\
% % 			MoSSo-Lossy (U)& Alg.~\ref{alg:dynamic} & $\UnweightedSummaryGraph$  & N/A & Clusters \cite{buehrer2008scalable} & Eq.~\eqref{eq:SSumMUnweightedObjective} & N/A\\
% 			MoSSo-Lossy (W)& Alg.~\ref{alg:dynamic} & $\UnweightedSummaryGraph$ & N/A & Cluster \cite{buehrer2008scalable} & Eq.~\eqref{eq:SSumMObjective} & N/A\\
% 			MoSSo-Lossy (U)& Alg.~\ref{alg:dynamic} & $\UnweightedSummaryGraph$  & N/A & Cluster \cite{buehrer2008scalable} & Eq.~\eqref{eq:SSumMUnweightedObjective} & N/A\\
% 			\bottomrule
% 		\end{tabular}
		\begin{tabular}{l|c|c|c|c|c|c|c}
			\toprule 
			Algorithm & Outline & \ $\SummaryGraph$ \ & $T$ & $size()$ & $groups()$ & $loss()$ & $sparsify?()$ \\
			\midrule
			k-Grass (W) & Alg.~\ref{alg:static} & $\WeightedSummaryGraph$ & Infinite &  $|S|$ & $\{S\}$ & Eq.~\eqref{eq:ReconstructionError} & False \\
			k-Grass (U) &  Alg.~\ref{alg:static} &$\UnweightedSummaryGraph$ & Infinite & $|S|$ & $\{S\}$ & Eq.~\eqref{eq:ReconstructionError} & False  \\
			SSumM (W)& Alg.~\ref{alg:static} & $\WeightedSummaryGraph$ & Finite & Eq.~\eqref{eq:size:weighted} & Clusters \cite{buehrer2008scalable} & Eq.~\eqref{eq:SSumMObjective} & True  \\
			SSumM (U) & Alg.~\ref{alg:static} & $\UnweightedSummaryGraph$ & Finite &  Eq.~\eqref{eq:size:unweighted} & Clusters \cite{buehrer2008scalable} & Eq.~\eqref{eq:SSumMUnweightedObjective} & True  \\
% 			MoSSo-Lossy (W)& Alg.~\ref{alg:dynamic} & $\UnweightedSummaryGraph$ & N/A & Clusters \cite{buehrer2008scalable} & Eq.~\eqref{eq:SSumMObjective} & N/A\\
% 			MoSSo-Lossy (U)& Alg.~\ref{alg:dynamic} & $\UnweightedSummaryGraph$  & N/A & Clusters \cite{buehrer2008scalable} & Eq.~\eqref{eq:SSumMUnweightedObjective} & N/A\\
			MoSSo-Lossy (W)& Alg.~\ref{alg:dynamic} & $\WeightedSummaryGraph$ & N/A & N/A & Clusters \cite{buehrer2008scalable} & Eq.~\eqref{eq:SSumMObjective} & N/A  \\
			MoSSo-Lossy (U)& Alg.~\ref{alg:dynamic} & $\UnweightedSummaryGraph$ & N/A & N/A & Clusters \cite{buehrer2008scalable} & Eq.~\eqref{eq:SSumMUnweightedObjective} & N/A \\
			\bottomrule
		\end{tabular}
	\end{center}
\begin{algorithm}[H]
    \small
    % \caption{\kGrass,\SSumM}\label{alg:static}
    \caption{Batch computation of a summary graph}\label{alg:static}
    \KwInput{(1) input graph $G$, (2) budget $k$, and (3) \# iters: $T$}
    \KwOutput{summary graph $\SummaryGraph$}
    \SetAlgoLined
        % \normalfont{initialize} \red{$\WeightedSummaryGraph$} \normalfont{or} \blue{$\UnweightedSummaryGraph$} \\
        % \normalfont{initialize} $summaryGraph$ \\
        \normalfont{initialize} $\SummaryGraph$; $t\leftarrow 1$ \\
        \While{$size(\SummaryGraph)$$> k$ \normalfont{and} $t<T$}{
            % \normalfont{get the best pair} $\{A, B\}$\\
            % \normalfont{to minimize} Eq~.\red{(\ref{eq:SSumMObjective})},\blue{(\ref{eq:SSumMUnweightedObjective})}\\
            % \normalfont{merge(}$A$, $B$\normalfont{)}
            
            $C \leftarrow$ $groups()$; $t\leftarrow t+1$\\ %Candidate Sets
            % \normalfont{merge a pair} $\{A, B\}$ among ${C \choose 2}$\\
             \For{$\mathbf{each}$ $C_{i}\in C$}{
                 \normalfont{merge one or more pairs} within $C_{i}$ 
                 \normalfont{to minimize} $loss()$ 
                 \\
             }
            %\normalfont{\textbf{if}} \blue{$add?()$} \normalfont{\textbf{then}}  \\
            % \normalfont{to minimize} $Loss()$ \\
           % \blue{merge\normalfont{(}$C$\normalfont{)}} \\
            % \normalfont{\textbf{for each} $S_{i} \in C$ \textbf{do} merge pairs among pairs of supernodes in $S_{i}$} \\
            
            % \normalfont{merge(}$A$, $B$\normalfont{)}
        }
        \textbf{if} $sparsify?()$ \textbf{then}
            % \normalfont{sparsify \red{$\WeightedSummaryGraph$} \normalfont{or} \blue{$\UnweightedSummaryGraph$}}
            % \normalfont{sparsify} $summaryGraph$ \\
            \normalfont{sparsify} $\SummaryGraph$ until $size(\SummaryGraph)$$\leq k$ \\
        \normalfont{\textbf{return}}  $\SummaryGraph$
\end{algorithm}
\begin{algorithm}[H]
    \small
    \caption{\small Incremental update of a summary graph}\label{alg:dynamic}
    \KwInput{(1) summary graph $\SummaryGraph$ and (2) change in $\{$\normalfont{src, dst}$\}$}
    \KwOutput{updated $\SummaryGraph$}
    \SetAlgoLined
        $C \leftarrow$ $groups()$; \\
        \For{$\mathbf{each}$ $u\in$ \{\normalfont{src, dst}\}}{
            % {-0.3mm}
            $\hat{N}_{u}\leftarrow$ \normalfont{sample neighbors of $u$} \\
            % {-0.3mm}
            \For{$\mathbf{each}$ $w\in \hat{N}_{u}$}{
                $P\leftarrow$ $C'\in C$ where $w\in C'$\\
                % sample random $v \in N_{u}\cup C_{w}$ \\
                % {-0.3mm}
                $v\leftarrow$ draw one in $\hat{N}_{u}\cap P$\\%\footnote{group with $w$}\\
                % \footnote{$candidates(w)$ denotes a set of subnodes having similar connectivity with $w$}
                %$v\leftarrow$ \normalfont{Uniform(}$C$\normalfont{)}\\
                % sample random $v \in N^{rand}_{u}\cup \blue{candidates(w)}$ \\
                % \normalfont{move} $w$ \normalfont{to minimize} \blue{$loss()$} \\    
                \textbf{if} $loss()$ \normalfont{drops} \textbf{then}
                    \normalfont{move} $w$ \normalfont{to} $S_{v}$\\
            }

        }
        \normalfont{\textbf{return}} $\SummaryGraph$
\end{algorithm}

\end{table}

\subsection{Weighted Graph Summarization Algorithms}
\label{sec:algo:weighted}

We introduce three searching algorithms for finding a weighted summary graph $\WeightedSummaryGraph = (S, P, \WeightFunction)$ of the input graph $G$.
% \red{We introduce three searching algorithms for finding $\WeightedSummaryGraph = (S, P, \WeightFunction)$ of $G$.}
See Algorithms~\ref{alg:static} and \ref{alg:dynamic} for their outlines and Table~\ref{tab:algorithmDetail} for details.

% \red{We outline the skeleton codes for the three algorithms in Alg~\ref{alg:static} and \ref{alg:dynamic}, and provide details about the codes colored according to algorithms in Table~\ref{tab:algorithmDetail}.}

\smallsection{\kGrass.}
\kGrass \cite{lefevre2010grass} first initializes the set $S$ of supernodes so that each subnode forms a singleton supernode. Then, it repeats greedily merging a supernode pair until $|S|$ reaches the target number (i.e., the given constraint). Specifically, in each step, among all supernode pairs, \kGrass merges a pair whose merger increases  Eq.~\eqref{eq:ReconstructionError} least. During the entire process, \kGrass creates a superedge between each supernode pair $A$ and $B$ (i.e., $\{A,B\}\in P$) if and only if $\EAB>0$.
% \kGrass \cite{lefevre2010grass} first initializes the set $S$ of supernodes so that each subnode forms a singleton supernode.
% Then, it repeats greedily merging a supernode pair until $|S|$ reaches the target number (i.e., the given constraint). Specifically, in each step, among $candidates()$, \kGrass merges a pair whose merger increases  $loss()$ least.
% During the entire process, \kGrass creates a superedge between each supernode pair $A$ and $B$ (i.e., $\{A,B\}\in P$) if and only if $\EAB>0$.

\smallsection{\SSumM.}
\SSumM \cite{lee2020ssumm} initializes $S$ as in \kGrass. Then, \SSumM divides $S$ into disjoint groups of supernodes with similar connectivity to find pairs to be merged efficiently. After that, in each group, \SSumM repeats merging a pair of supernodes whose merger decreases Eq.~\eqref{eq:SSumMObjective} most. 
% \SSumM \cite{lee2020ssumm} initializes $S$ as in \kGrass. Then, \SSumM divides $S$ into disjoint groups of supernodes with similar connectivity to find pairs to be merged efficiently. After that, in each group of $candidates()$, \SSumM repeats merging a pair of supernodes whose merger increases $loss()$ least. 

\begin{equation}
size_{bits}(\WeightedSummaryGraph) + \sum_{\{A, B\}\in P}\PIAB\cdot \mathcal{H}(\frac{\EAB}{\PIAB})
    + \sum_{\{A, B\}\notin P} 2\EAB \log_{2}|V|,
\label{eq:SSumMObjective}
\end{equation}

\noindent where $\mathcal{H}(\cdot)$ is the entropy function.
Eq.~\eqref{eq:SSumMObjective} considers both the size of a summary graph and the reconstruction error. Specifically, the second term is the number of bits for exactly restoring the subedeges between supernodes that are joined by superedges, 
and the third term is that for the other subedges (see \cite{lee2020ssumm} for details). 
% \noindent where $\mathcal{H}(\cdot)$ is the entropy function.
% Eq.~\eqref{eq:SSumMObjective} considers both the size of a summary graph and the reconstruction error. Specifically, the second and third terms of Eq.~\eqref{eq:SSumMObjective} are the number of bits for exactly restoring subedeges between supernodes that are joined by superedges and the other subedges, respectively (see \cite{lee2020ssumm} for details). 
During the process, the superedge between each supernode pair exists only when it decreases Eq.~\eqref{eq:SSumMObjective}. 
If $size_{bits}(\WeightedSummaryGraph)$ (i.e., Eq.~\eqref{eq:size:weighted}) cannot satisfy the given constraint (i.e., the target size) within the given number of iterations, \SSumM sparsifies $\WeightedSummaryGraph$ greedily based on Eq.~\eqref{eq:ReconstructionError} to satisfy the constraint.

\smallsection{\MoSSoLossy.}
\MoSSoLossy is a lossy variant of \MoSSo \cite{ko2020incremental}, which is a lossless graph compression algorithm. While processing subedges incrementally, it updates $\WeightedSummaryGraph= (S, P, \WeightFunction)$. Specifically, for each subedge $\{u,v\}$, it samples a fixed number of neighbors of $u$ and $v$. Then, for each such neighbor $w$, \MoSSoLossy moves $w$ from $S_{w}$ to the supernode which another sampled subnode with similar connectivity belongs to if this change decreases Eq.~\eqref{eq:SSumMObjective}. As in \SSumM, for each pair of supernodes, a superedge joins them only when it decreases Eq.~\eqref{eq:SSumMObjective}.
% \MoSSoLossy is a lossy variant of \MoSSo \cite{ko2020incremental}, which is a lossless graph compression algorithm. While processing subedges sequentially, it updates $\WeightedSummaryGraph= (S, P, \WeightFunction)$. Specifically, for each subedge $\{u,v\}$, it samples a fixed number of neighbors of $u$ and $v$. Then, for each such neighbor $w$, \MoSSoLossy moves $w$ from $S_{w}$ to the supernode which another sampled subnode with similar connectivity belongs to if this change decreases $loss()$. 
% As in \SSumM, for each pair of supernodes, a superedge joins them only when it decreases $loss()$.

\subsection{Unweighted Graph Summarization Algorithms}
\label{sec:algo:unweighted}

% \red{We extend the above algorithms for obtaining $\UnweightedSummaryGraph = (S, P)$ of $G$.}
We extend the above algorithms for obtaining an unweighted summary graph $\UnweightedSummaryGraph = (S, P)$ of the input graph $G$.
%delete
The differences are highlighted in Table~\ref{tab:algorithmDetail} with the outlines in Algorithms~\ref{alg:static} and \ref{alg:dynamic}.

\smallsection{\kGrassUnweighted.}
This variant is different from \kGrass in that Eq.~\eqref{eq:unweightedAdjValue} is used, instead of Eq.~\eqref{eq:WeightedAdjValue}, when Eq.~\eqref{eq:ReconstructionError} is computed. Moreover, for each supernode pair, the superedge between them exists only when it decreases Eq.~\eqref{eq:ReconstructionError}. 
% This variant is different from \kGrass in that Eq.~\eqref{eq:unweightedAdjValue} is used, instead of Eq.~\eqref{eq:WeightedAdjValue}, when $loss()$ is computed. 
% Moreover, for each supernode pair, the superedge between them exists only when it decreases $loss()$. 

\smallsection{\SSumMUnweighted.}
Instead of Eq.~\eqref{eq:SSumMObjective} used in \SSumM,
this variant uses Eq.~\eqref{eq:SSumMUnweightedObjective}, whose second term is the number of bits for exactly restoring the subedges between supernodes that are joined by unweighted superedges. 
% This variant has the different points in $\SummaryGraph$, $size()$ and $loss()$ in Table~\ref{tab:Algorithms}.
% The variant use $loss()$ as Eq.~\eqref{eq:SSumMUnweightedObjective}, whose second term is the number of bits for exactly restoring the subedges between supernodes that are joined by unweighted superedges. 
% \begin{equation}
% size_{bits}(\UnweightedSummaryGraph)+\sum_{\mathclap{\{A, B\}\in P}}2\left(\PIAB-\EAB\right) \log_{2}|V| + \sum_{\mathclap{\{A, B\}\notin P}}2\EAB \log_{2}|V|.
% \label{eq:SSumMUnweightedObjective}
% \end{equation}
\begin{equation}
size_{bits}(\UnweightedSummaryGraph)+\sum_{\{A, B\}\in P}2\left(\PIAB-\EAB\right) \log_{2}|V| + \sum_{\{A, B\}\notin P}2\EAB \log_{2}|V|.
\label{eq:SSumMUnweightedObjective}
\end{equation}

\smallsection{\MoSSoLossyUnweighted.}
This variant uses Eq.~\eqref{eq:SSumMUnweightedObjective}, instead of  Eq.~\eqref{eq:SSumMObjective}, which is used in \MoSSoLossy. 
% This variant has differences in $loss()$ and $\SummaryGraph$. 

\section{Experiments}\label{sec:experiments}
%example table

We review our experiments for comparing weighted and unweighted graph summarization in five  aspects.
We describe the settings and then present the results.

\begin{table}[t!]
    \centering
	\begin{center}
		\caption{Summary of the eight real-world graphs used in the paper. They are obtained from emails (EE), collaborations (DB), co-purchases (A6), computer networks (SK), online social networks (LJ), and hyperlinks (WS, DP, and WL).}\label{tab:DatasetTable}
		\small
		\begin{tabular}{c|c|c||c|c|c}
			\toprule 
			\textbf{Name} & \textbf{\# Nodes} & \textbf{\# Edges}   & \textbf{Name} & \textbf{\# Nodes} & \textbf{\# Edges}  \\
			\midrule
			Email-Enron (EE) %\cite{klimt2004introducing} 
			& 36,692\ & 183,831 \ &
			\ DBLP (DB) %\cite{yang2015defining} 
			& 317,080\ & \ 1,049,866 \\
			Amazon-0601 (A6) %\cite{leskovec2007dynamics} 
			& 403,394\ & 2,443,408 \ &
			\ WebSmall (WS) %\cite{BoVWFI}     
			& 325,557\ & \ 2,738,969                 \\
			Skitter (SK) %\cite{leskovec2005graphs} 
			& 1,696,415\ & 11,095,298 \ &
			\ LiveJournal (LJ) %\cite{yang2015defining} 
			& 3,997,962\ & \ 34,681,189\\
			DBPedia (DP) %\cite{auer2007dbpedia,kunegis2013konect} 
			& 18,268,991\ & 126,890,209 \ &
			\ WebLarge (WL) %\cite{boldi2004webgraph} 
			& 18,483,186\ & \ 261,787,258\\
			\bottomrule 
		\end{tabular}
	\end{center}
\end{table}

\subsection{Experimental Settings}

\smallsection{Machines:}
We performed our experiments on a desktop with a 3.80GHZ Intel i7-10700K CPU and 64GB memory.

\smallsection{Datasets:}
We used the eight datasets summarized in Table~\ref{tab:DatasetTable}.

\smallsection{Search Algorithms:}
We used the six algorithms described in Sect.~\ref{sec:problem}. We implemented them commonly in OpenJDK 12 and set their target size to $\{0.1,0.2,\cdots,0.9\}$ of the size in the input graph. We fixed $T$ to $20$ in both versions of \SSumM.
% We excluded \cite{khan2015set,ko2020incremental,navlakha2008graph,shin2019sweg} from the comparison since they assume extra components (e.g., edge corrections) in addition to a summary graph.
We excluded \cite{ko2020incremental,navlakha2008graph,shin2019sweg} from the comparison since they assume extra components (e.g., edge corrections) in addition to a summary graph.

\smallsection{Evaluation Metric:}
Given the input graph $G=(V,E)$ and a summary graph $G'$ (i.e., $\WeightedReconstructedGraph$ or $\UnweightedSummaryGraph$), the compression ratio is defined in bits as $\frac{size_{bits}(G')}{2|E|\log_{2}|V|}$. %(see Eqs.~\eqref{eq:size:weighted} and \eqref{eq:size:unweighted} for $size_{bits}(G')$)

\subsection{Results}

\begin{figure*}[t!]
	\centering
	\includegraphics[width=0.7\linewidth]{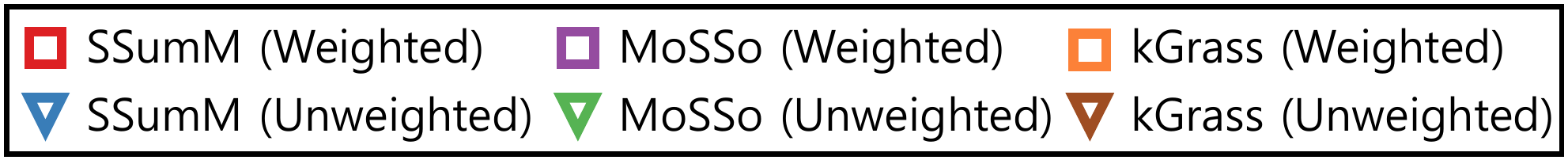} \\
	\textbf{L1 Reconstruction Error:} \hfill \ \ \ \\
	\subfigure[Email-Enron]{
		\label{fig:Reconstruction L1 EE}
		\includegraphics[width=0.22\textwidth]{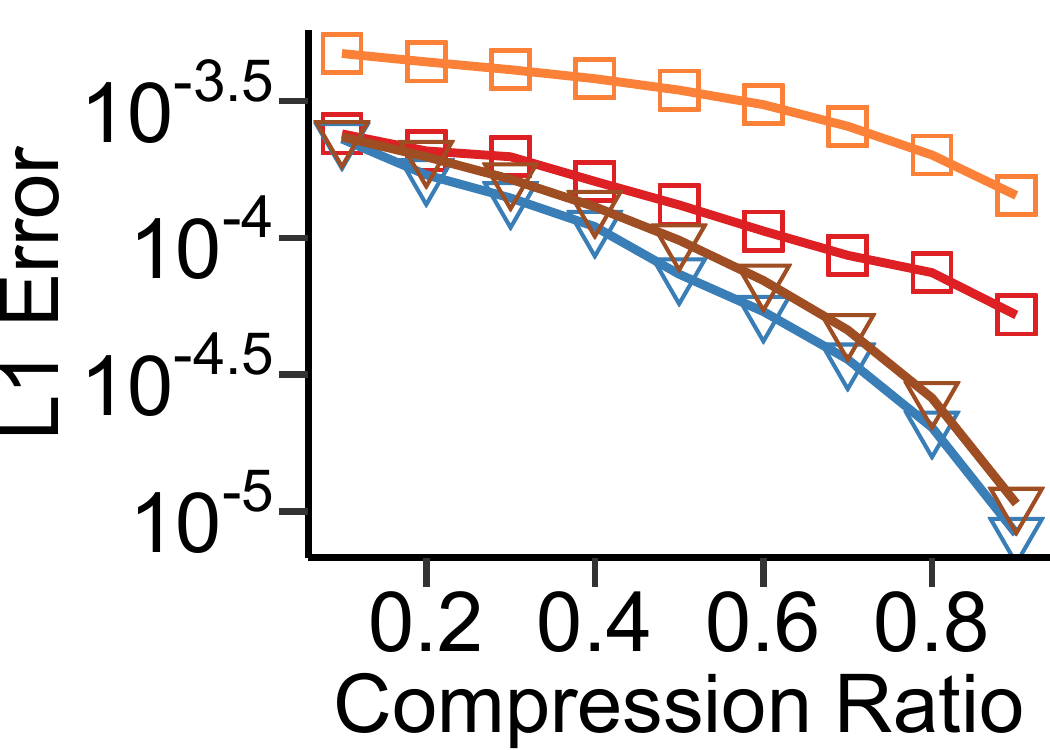}
	}
	\subfigure[DBLP]{
		\label{fig:Reconstruction L1 DB}
		\includegraphics[width=0.22\textwidth]{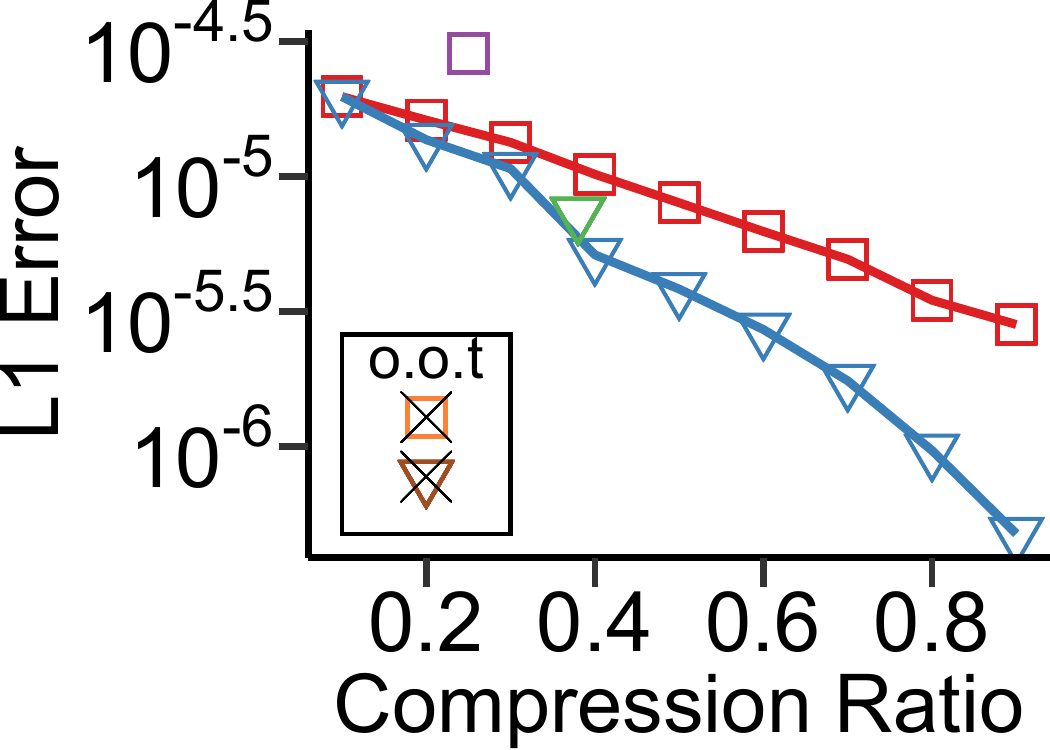}
	} 
	\subfigure[Amazon-0601]{
		\label{fig:Reconstruction L1 A6}
		\includegraphics[width=0.22\textwidth]{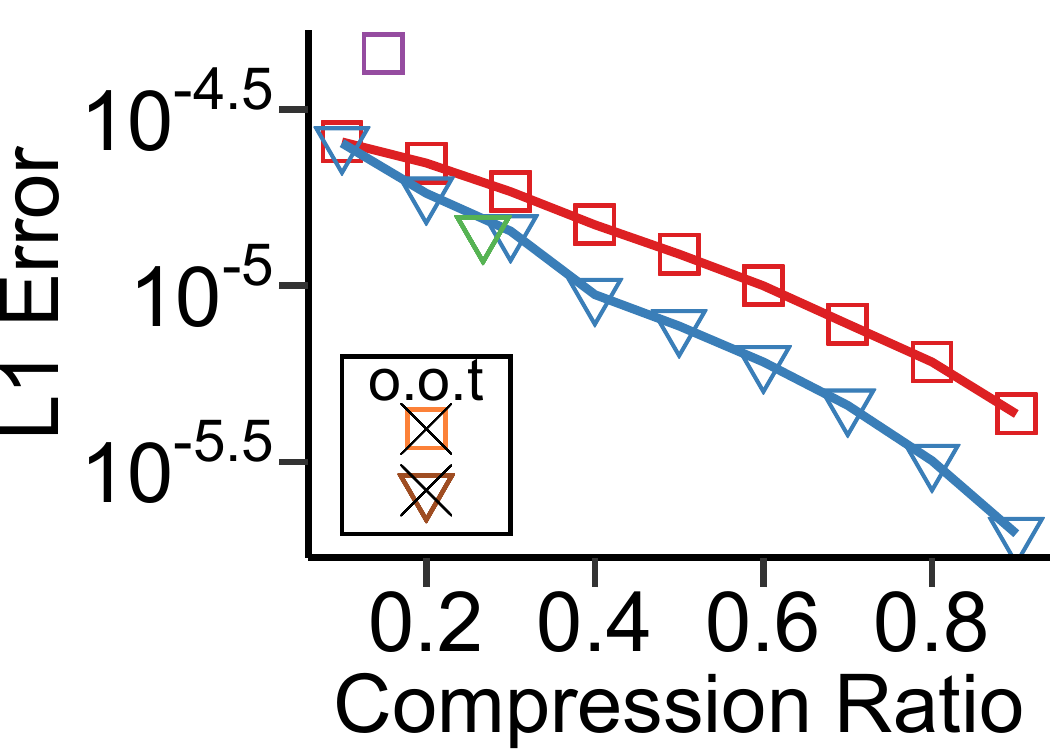}
	}
	\subfigure[WebSmall]{
		\label{fig:Reconstruction L1 C2}
		\includegraphics[width=0.22\textwidth]{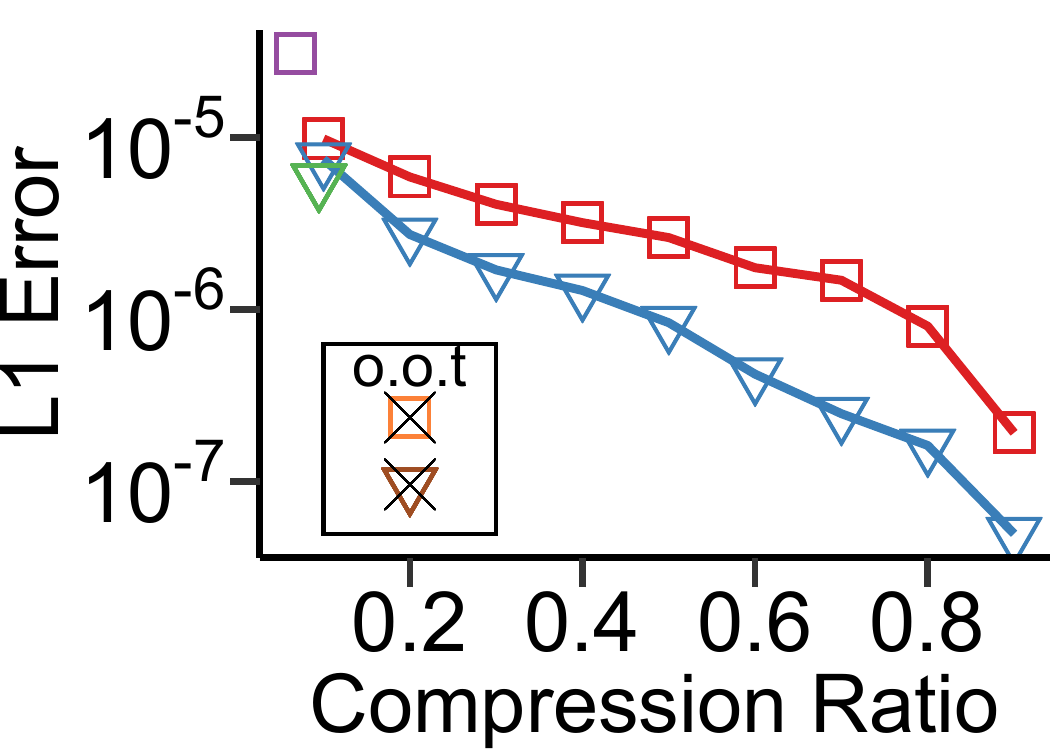}
	} \\
	\subfigure[Skitter]{
		\label{fig:Reconstruction L1 SK}
		\includegraphics[width=0.22\textwidth]{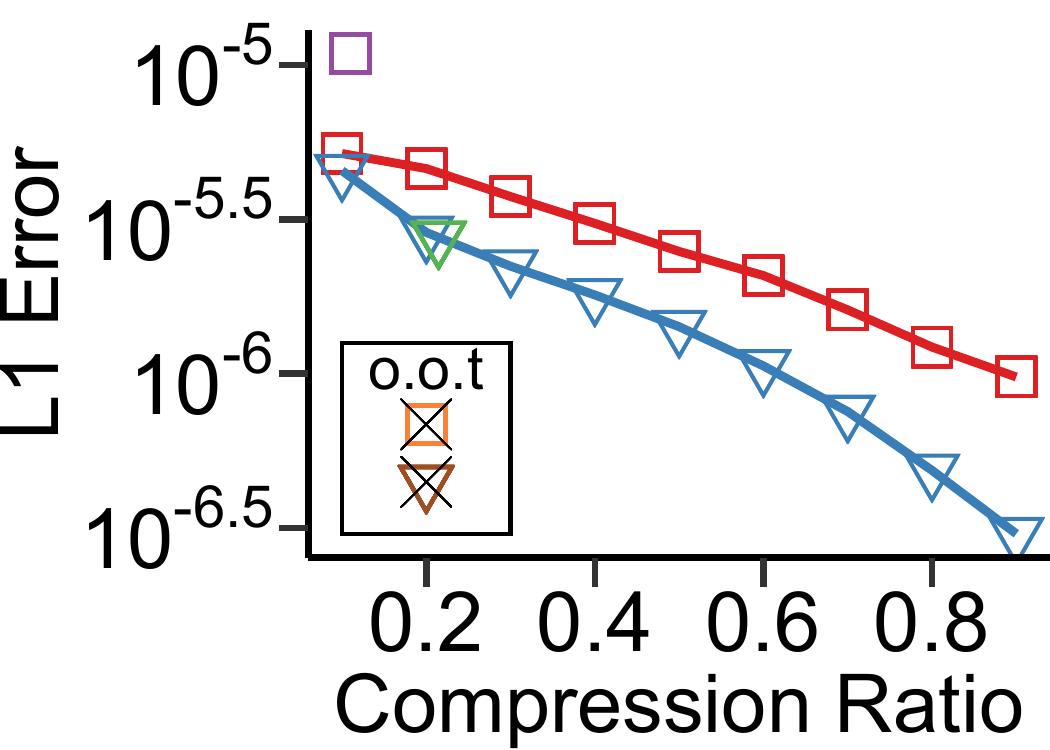}
	}
	\subfigure[LiveJournal]{
		\label{fig:Reconstruction L1 LJ}
		\includegraphics[width=0.22\textwidth]{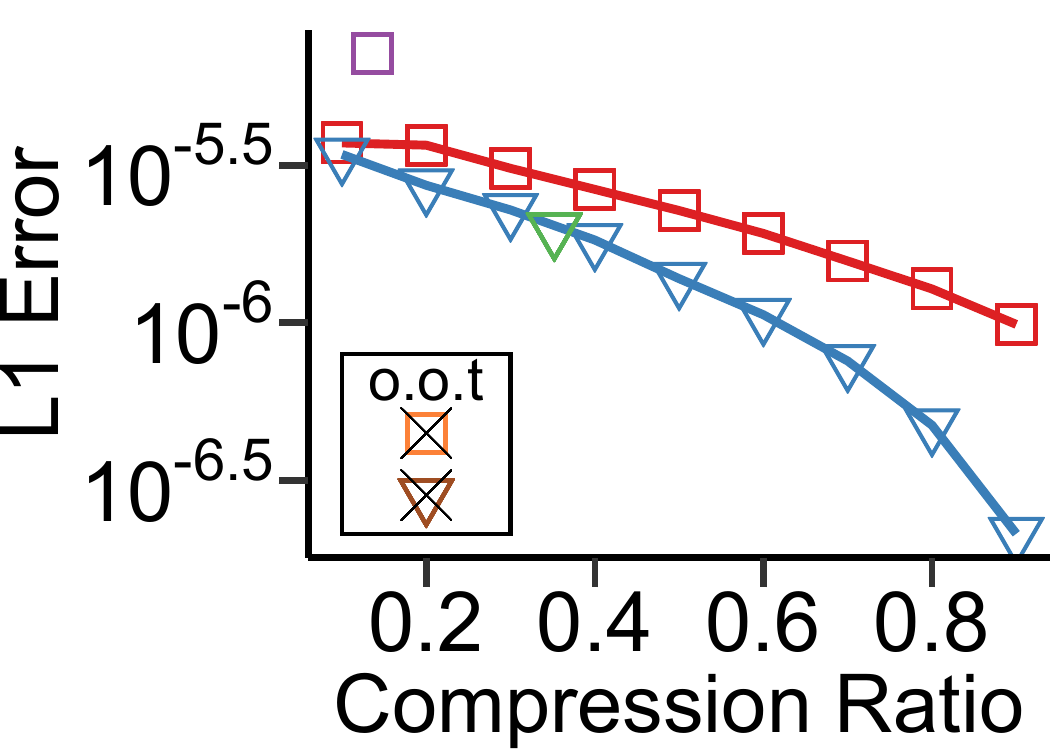}
	}
	\subfigure[DBPedia]{
		\label{fig:Reconstruction L1 C2}
		\includegraphics[width=0.22\textwidth]{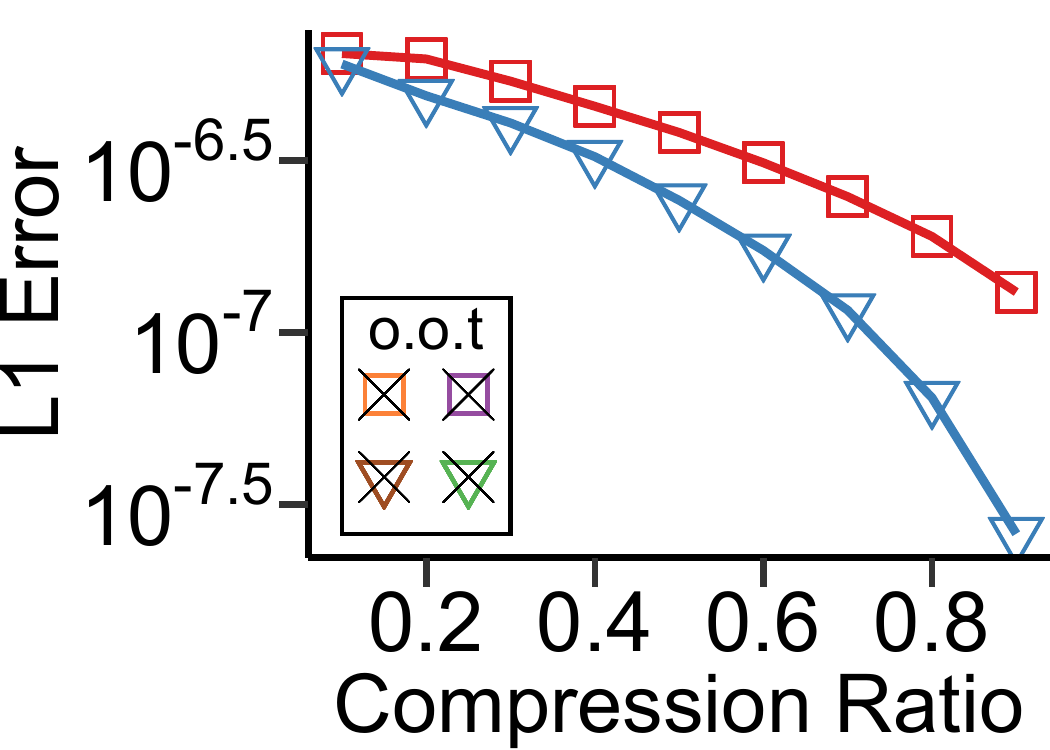}
	}  
	\subfigure[WebLarge]{
		\label{fig:Reconstruction L1 C2}
		\includegraphics[width=0.22\textwidth]{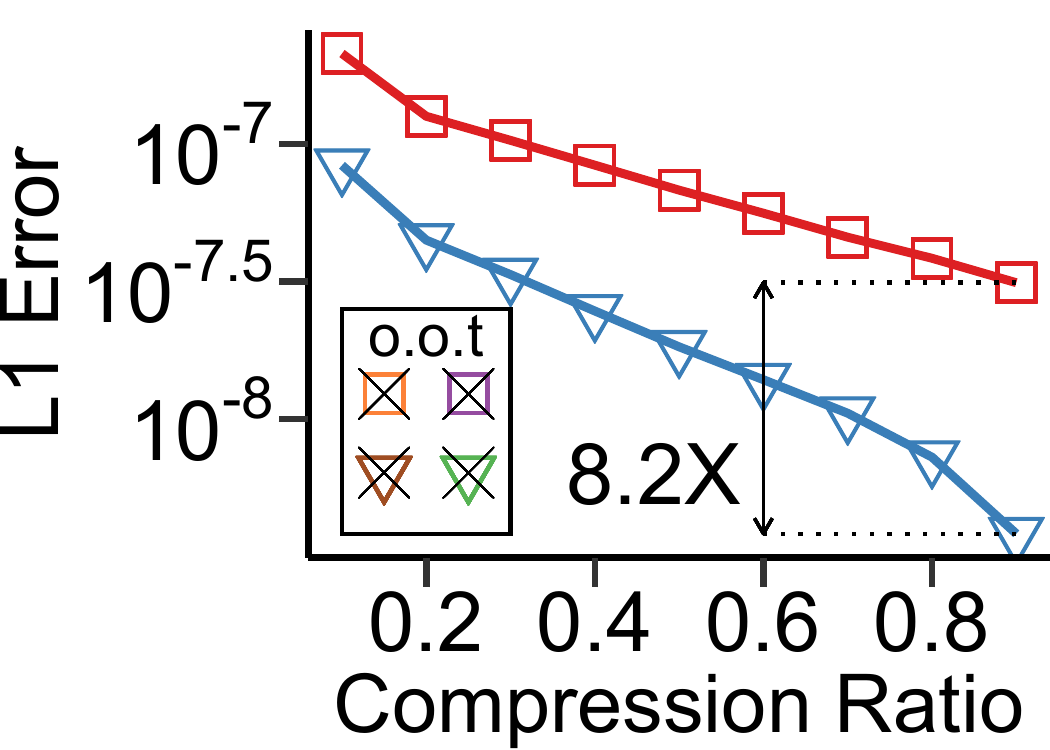}
	}\\
	\textbf{L2 Reconstruction Error:} \hfill \ \ \ \\
	\subfigure[Email-Enron]{
		\label{fig:Reconstruction L2 EE}
		\includegraphics[width=0.22\textwidth]{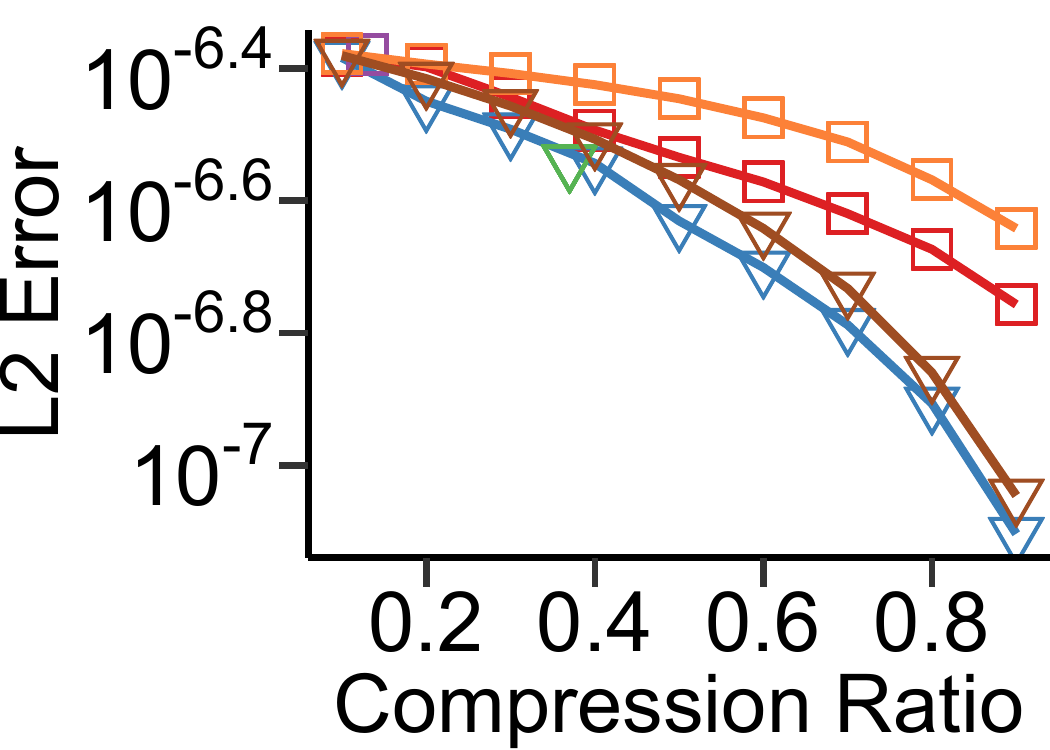}
	}
	\subfigure[DBLP]{
		\label{fig:Reconstruction L2 DB}
		\includegraphics[width=0.22\textwidth]{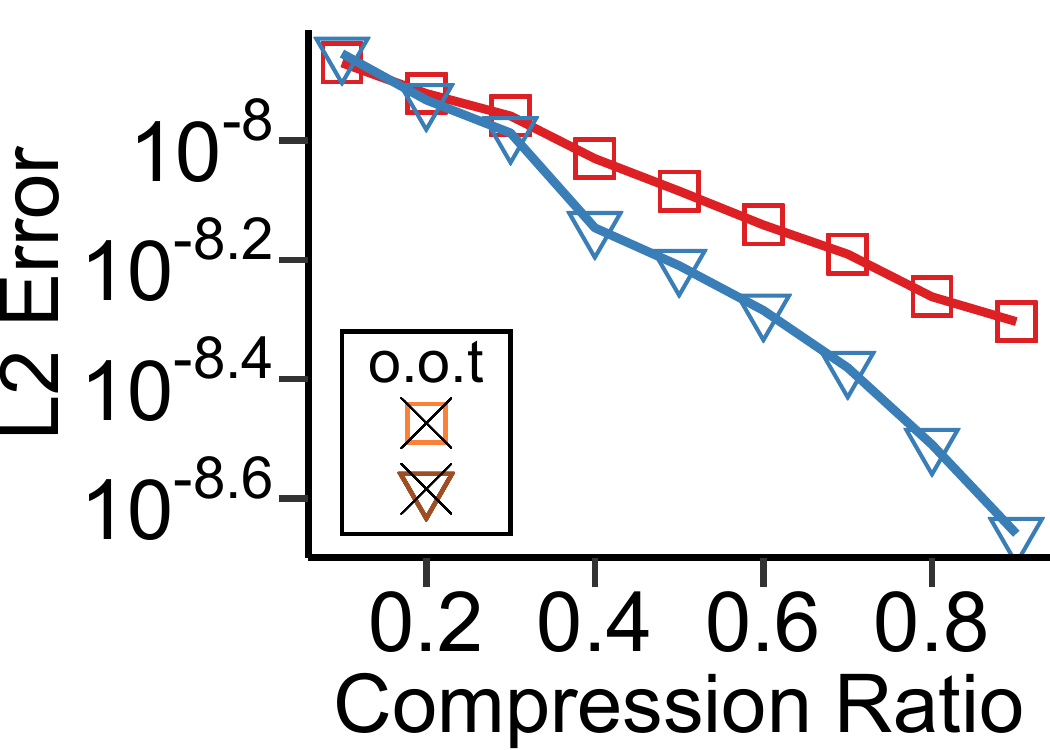}
	} 
	\subfigure[Amazon-0601]{
		\label{fig:Reconstruction L2 A6}
		\includegraphics[width=0.22\textwidth]{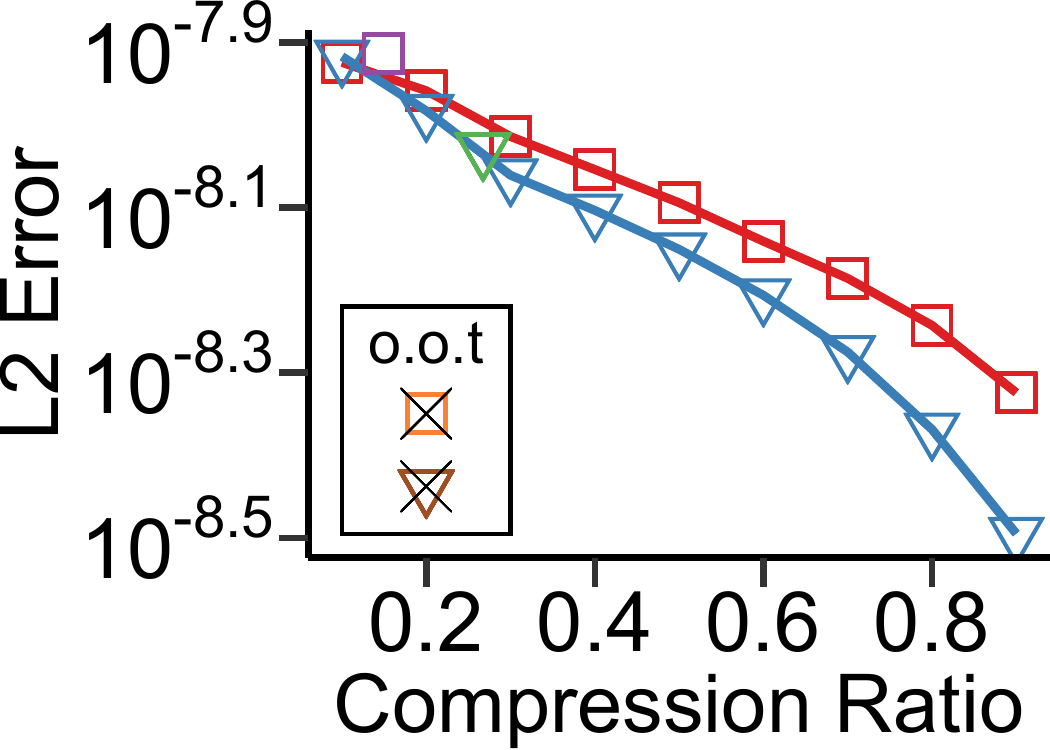}
	}
	\subfigure[WebSmall]{
		\label{fig:Reconstruction L2 C2}
		\includegraphics[width=0.22\textwidth]{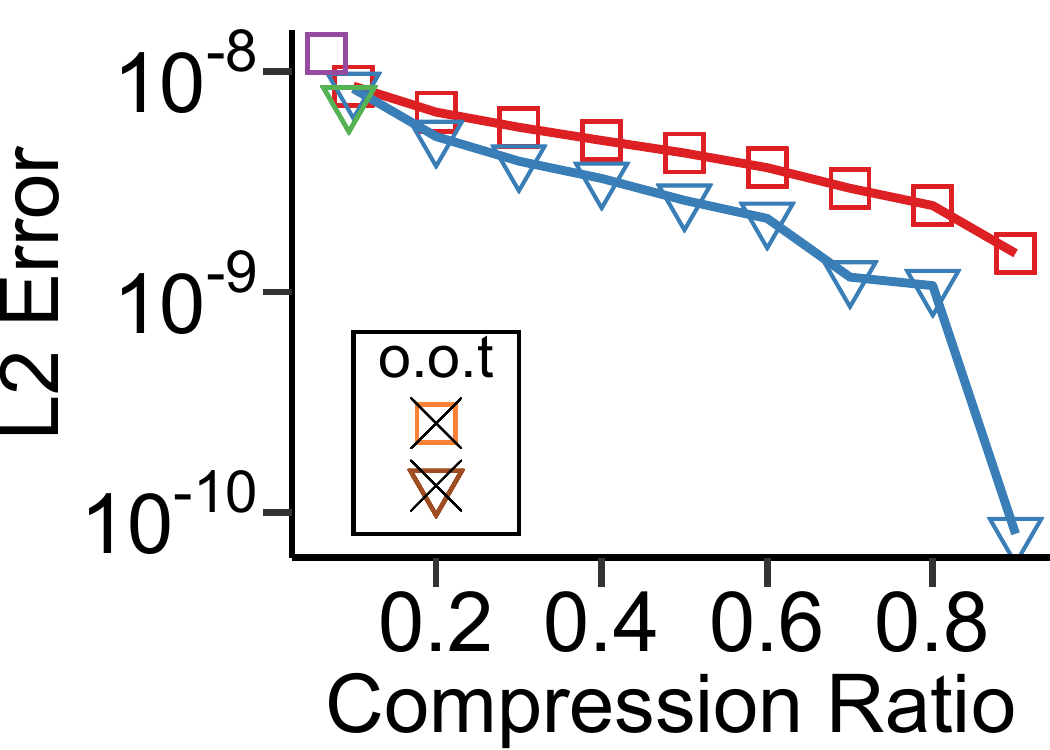}
	}  
	\\
	\subfigure[Skitter]{
		\label{fig:Reconstruction L2 SK}
		\includegraphics[width=0.22\textwidth]{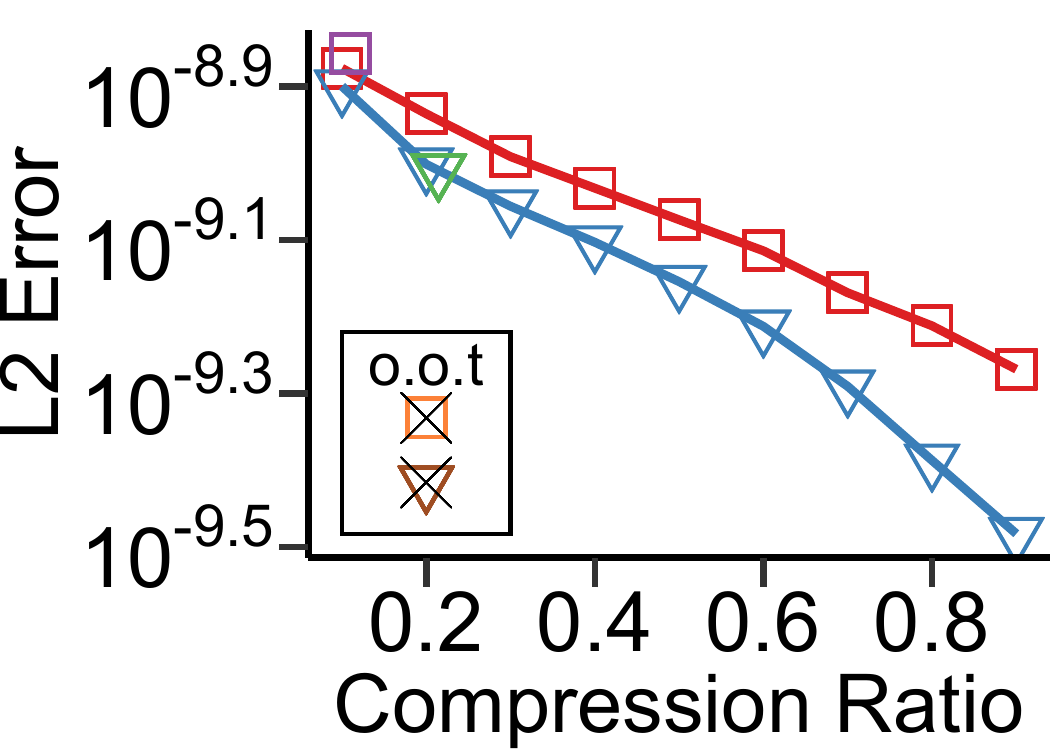}
	}
	\subfigure[LiveJournal]{
		\label{fig:Reconstruction L2 LJ}
		\includegraphics[width=0.22\textwidth]{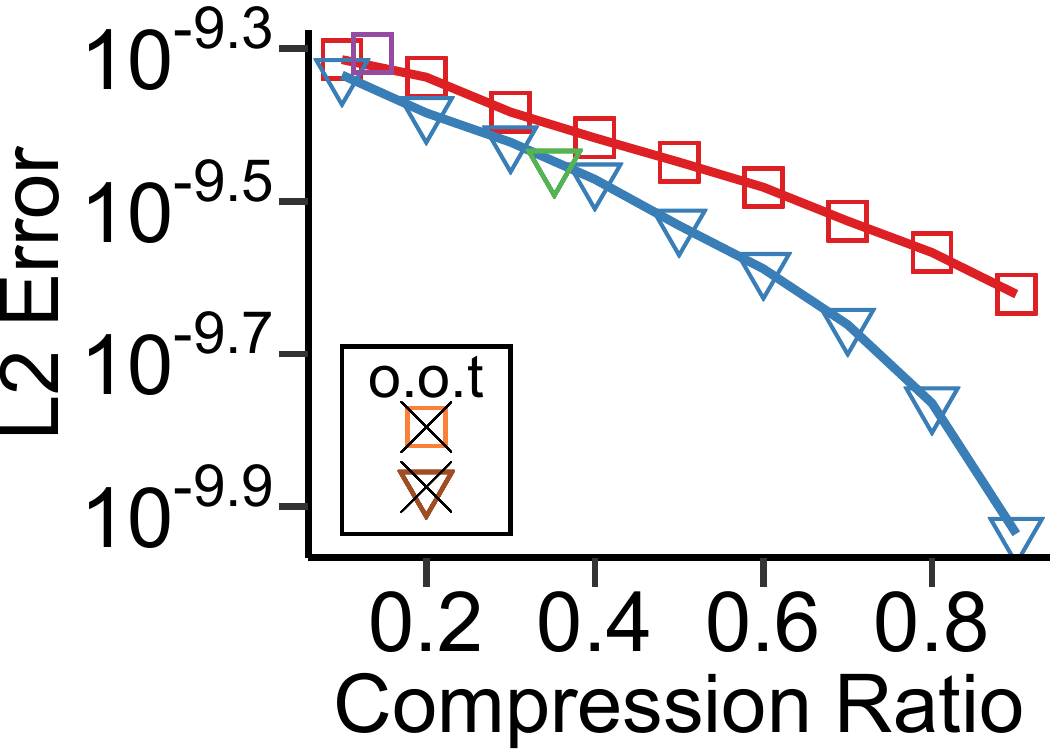}
	} 
	\subfigure[DBPedia]{
		\label{fig:Reconstruction L2 DP}
		\includegraphics[width=0.22\textwidth]{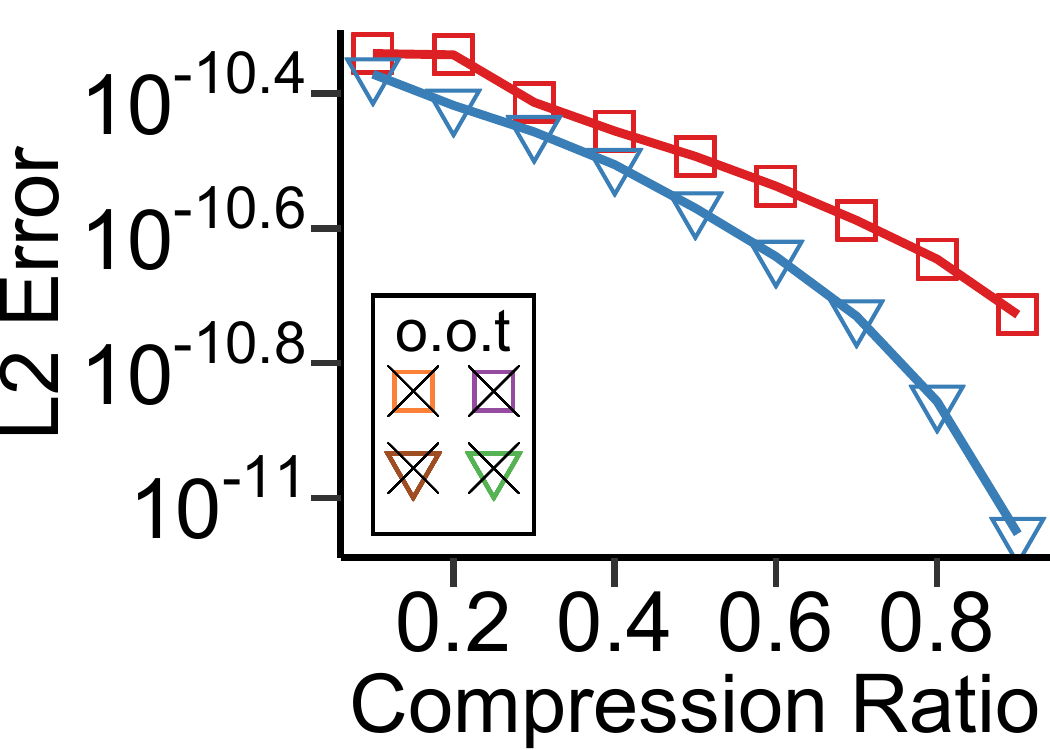}
	}  
	\subfigure[WebLarge]{
		\label{fig:Reconstruction L2 UK}
		\includegraphics[width=0.22\textwidth]{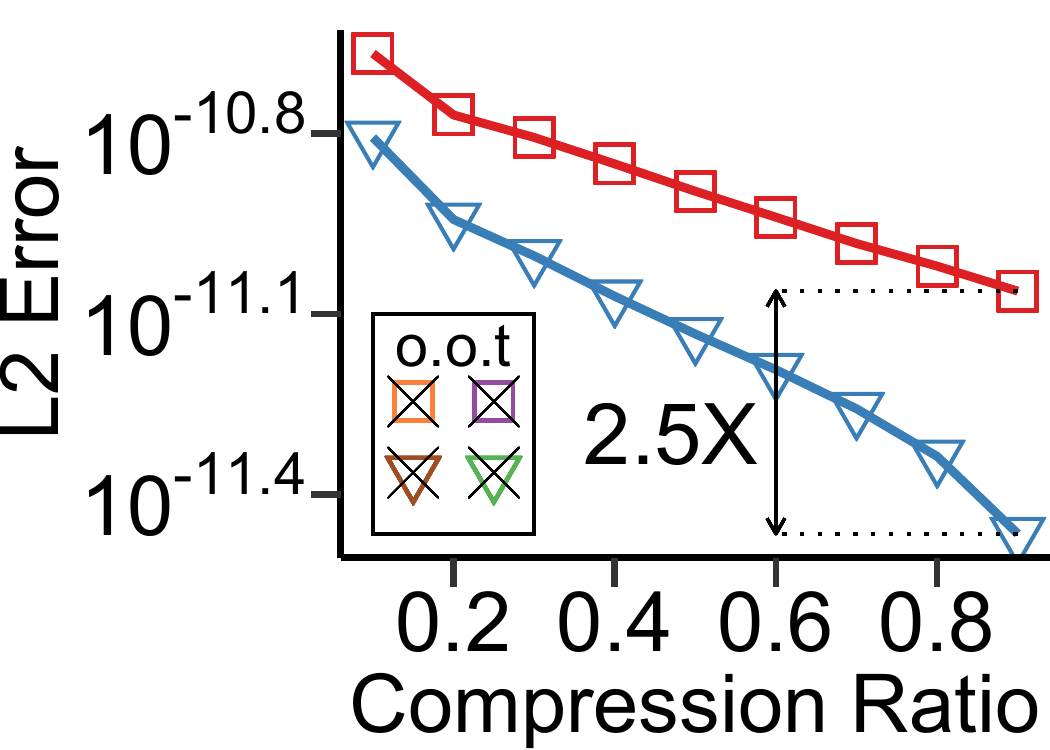}
	} \\
    \caption{\label{fig:ExperimentsResult_L1L2}
    The reconstruction error is significantly lower in unweighted graph summarization than in weighted summarization. \textbf{o.o.t.}: out of time ($\geq$ 48 hours).
    % Unweighted ones are consistently better than weighted ones in all considered aspects.
	}
\end{figure*}

\smallsection{Reconstruction Error:} The $L_{1}$ and $L_{2}$ reconstruction error (i.e., $p=\{1,2\}$ in Eq.~\eqref{eq:ReconstructionError}) is compared in
Fig.~\ref{fig:ExperimentsResult_L1L2}.
Unweighted summary graphs described the input graph more accurately (specifically, up to $8.2\times$ when comparing \SSumM and its variant) than weighted ones, when compression ratios were the same.

% \begin{figure*}[t!]
% 	\centering
% % 	\includegraphics[width=1\linewidth]{./ExperimentResult/legend.png} 
% 	\subfigure[Email-Enron]{
% 		\label{fig:Reconstruction L2 EE}
% 		\includegraphics[width=0.22\textwidth]{ExperimentResult/L2RE/EE_RE.pdf}
% 	}
% 	\subfigure[DBLP]{
% 		\label{fig:Reconstruction L2 DB}
% 		\includegraphics[width=0.22\textwidth]{ExperimentResult/L2RE/DB_RE.pdf}
% 	} 
% 	\subfigure[Amazon-0601]{
% 		\label{fig:Reconstruction L2 A6}
% 		\includegraphics[width=0.22\textwidth]{ExperimentResult/L2RE/A6_RE.pdf}
% 	}
% 	\subfigure[WebSmall]{
% 		\label{fig:Reconstruction L2 C2}
% 		\includegraphics[width=0.22\textwidth]{ExperimentResult/L2RE/C2_RE.pdf}
% 	}  
% 	\\
% 	\subfigure[Skitter]{
% 		\label{fig:Reconstruction L2 SK}
% 		\includegraphics[width=0.22\textwidth]{ExperimentResult/L2RE/SK_RE.pdf}
% 	}
% 	\subfigure[LiveJournal]{
% 		\label{fig:Reconstruction L2 LJ}
% 		\includegraphics[width=0.22\textwidth]{ExperimentResult/L2RE/LJ_RE.pdf}
% 	} 
% 	\subfigure[DBPedia]{
% 		\label{fig:Reconstruction L2 DP}
% 		\includegraphics[width=0.22\textwidth]{ExperimentResult/L2RE/DP_RE.pdf}
% 	}  
% 	\subfigure[WebLarge]{
% 		\label{fig:Reconstruction L2 UK}
% 		\includegraphics[width=0.22\textwidth]{ExperimentResult/L2RE/UK_RE.pdf}
% 	}  
%     \caption{
%     Comparison of weighted and unweighted summary graphs in terms of $L_{2}$ reconstruction error. \textbf{o.o.t.}: out of time ($\geq$ 48 hours). 
%     % Unweighted ones are consistently better than weighted ones in all considered aspects.
%     \label{fig:ExperimentsResult_L2RE}
% 	}
% \end{figure*}

\begin{figure*}[t!]
	\centering
	\includegraphics[width=0.7\linewidth]{./ExperimentResult/legend.pdf} \\
	\textbf{Error in Node Importance (Spec., PageRank  Scores \cite{page1999pagerank}):} \hfill \ \ \ \\
	\subfigure[Email-Enron]{
		\label{fig:PageRank L1 EE}
		\includegraphics[width=0.22\textwidth]{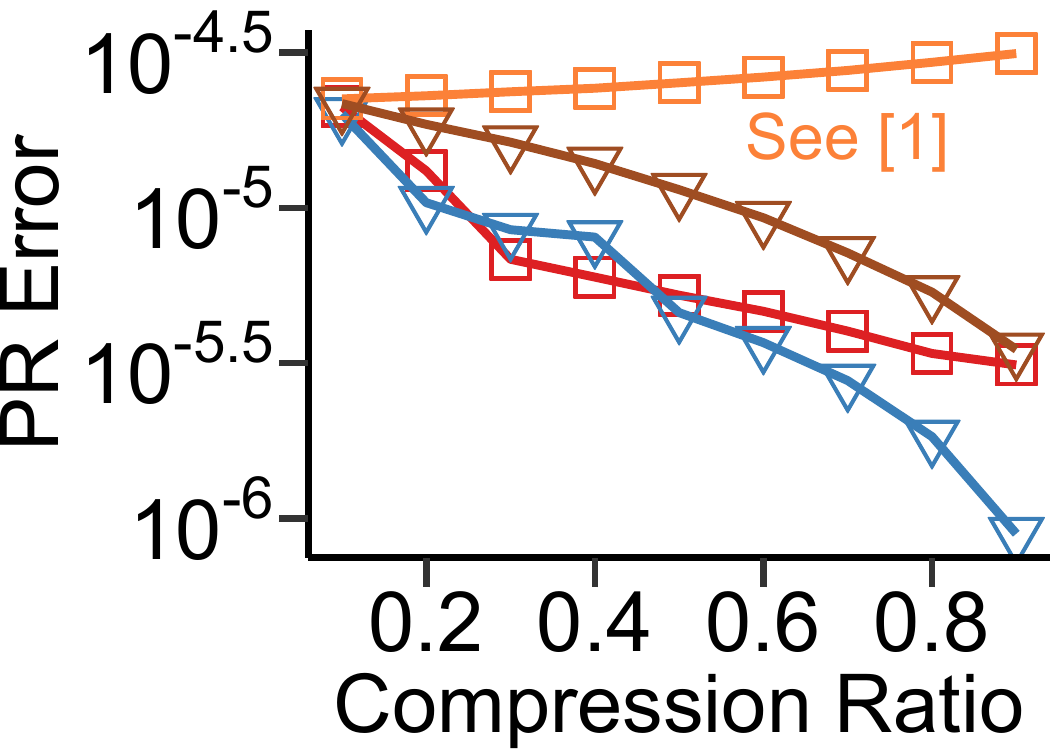}
	}
	\subfigure[DBLP]{
		\label{fig:PageRank L1 DB}
		\includegraphics[width=0.22\textwidth]{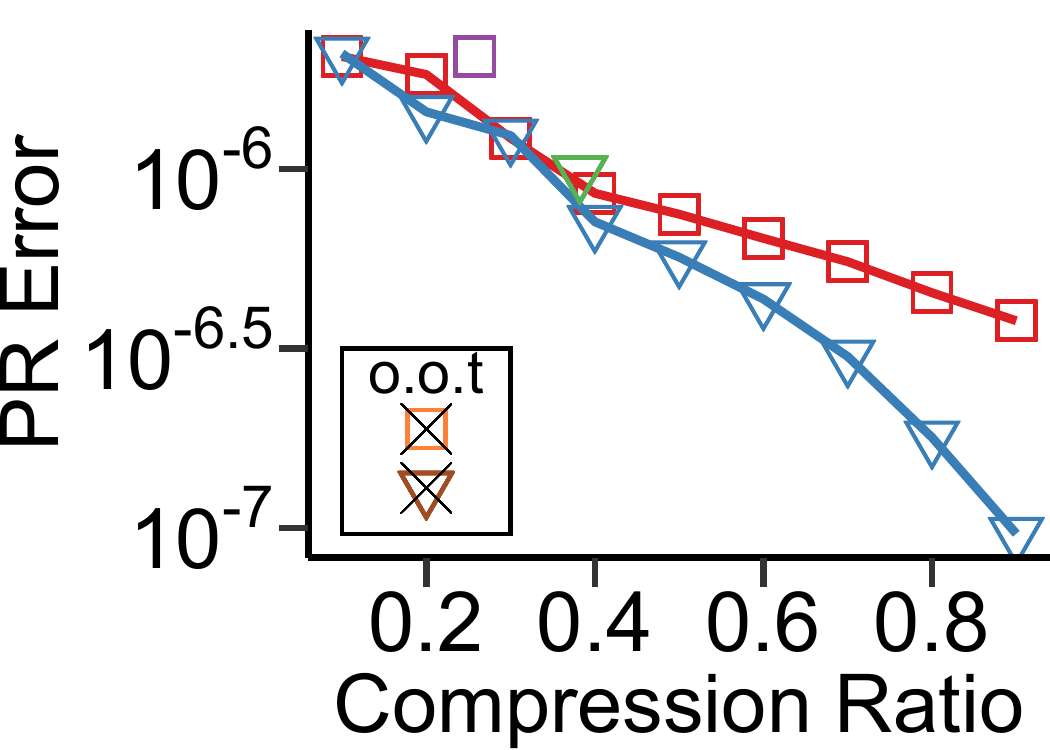}
	} 
	\subfigure[Amazon-0601]{
		\label{fig:PageRank L1 A6}
		\includegraphics[width=0.22\textwidth]{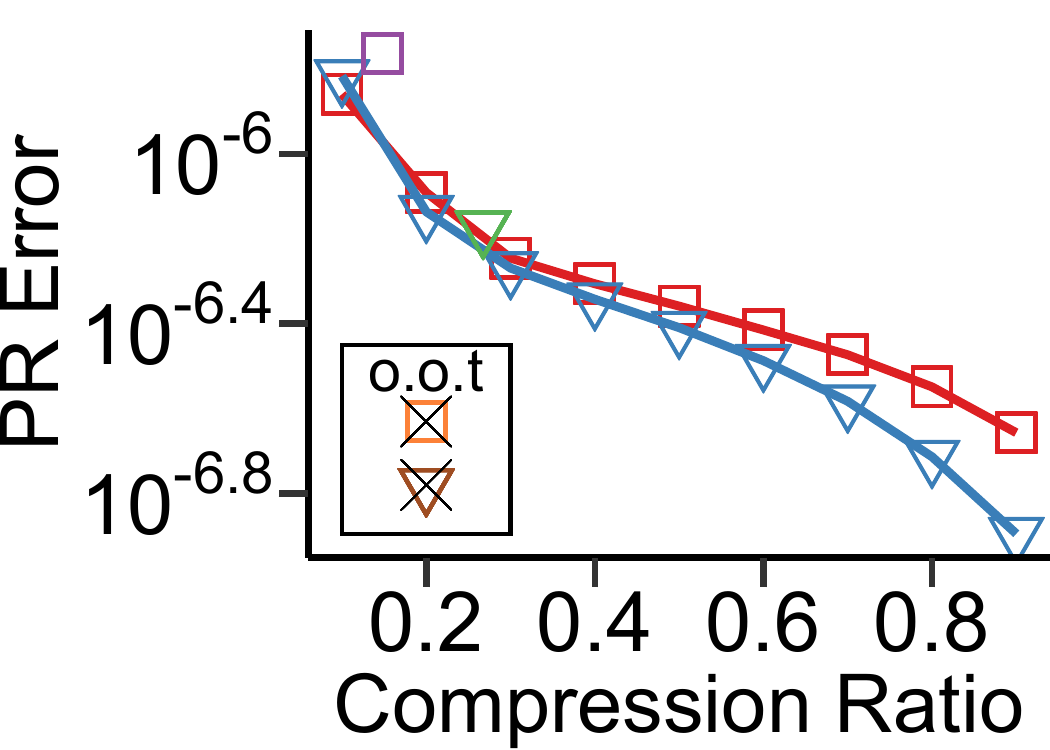}
	}
	\subfigure[WebSmall]{
		\label{fig:PageRank L1 C2}
		\includegraphics[width=0.22\textwidth]{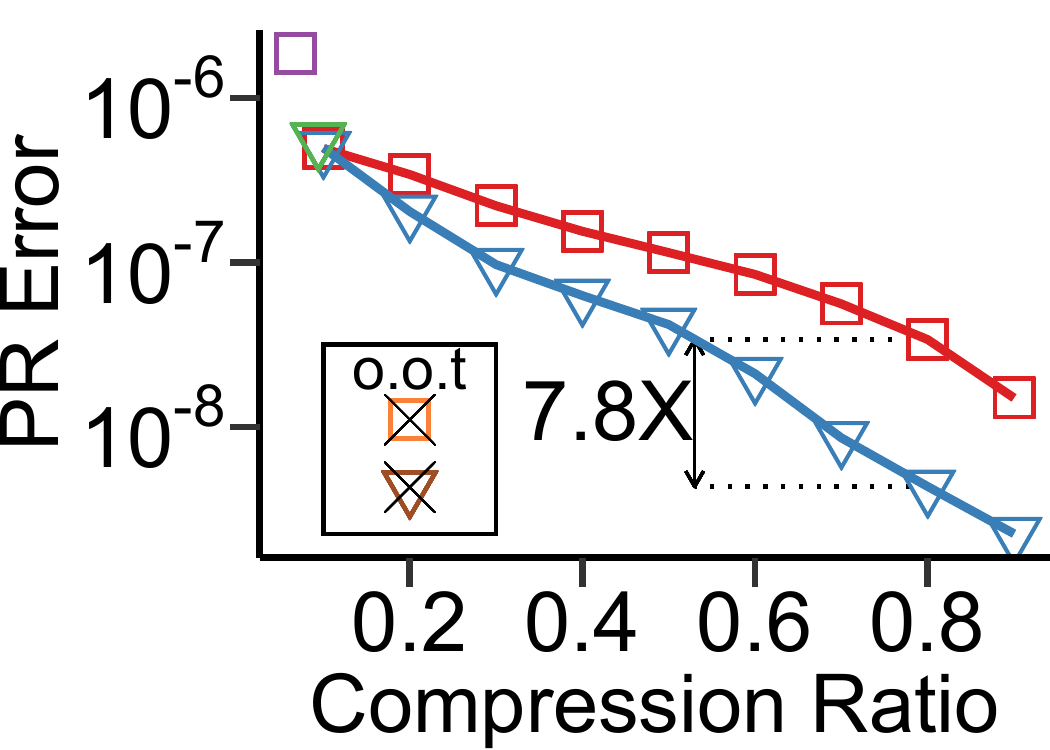}
	}
	\\
	\subfigure[Skitter]{
		\label{fig:PageRank L1 SK}
		\includegraphics[width=0.22\textwidth]{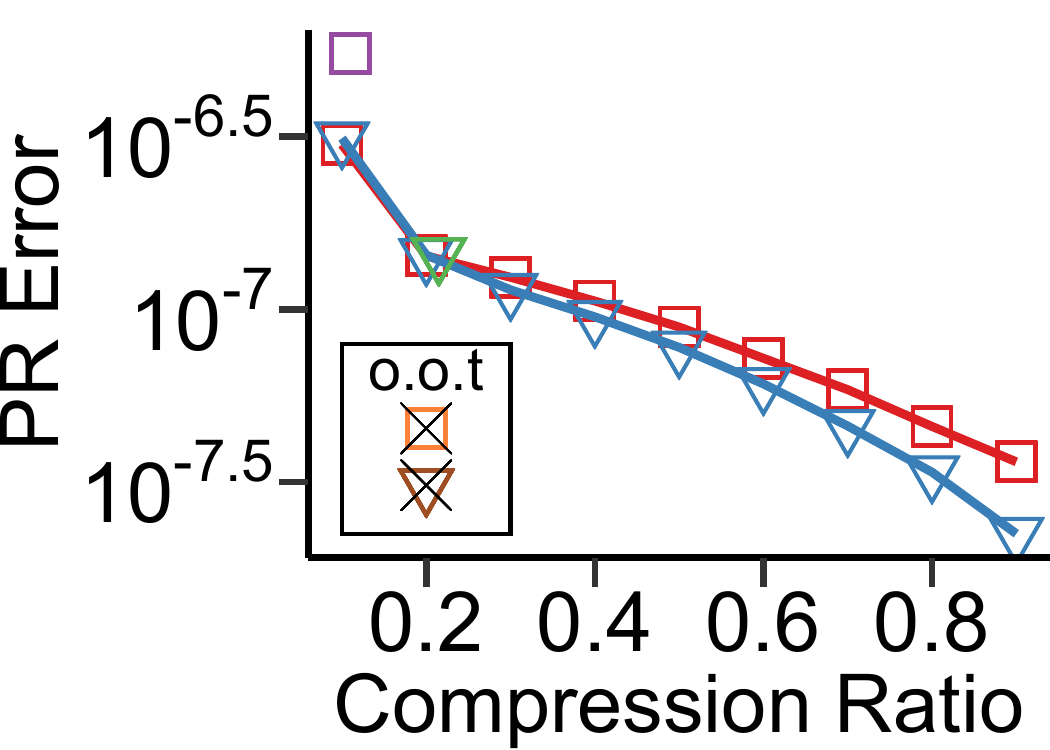}
	}
	\subfigure[LiveJournal]{
		\label{fig:PageRank L1 LJ}
		\includegraphics[width=0.22\textwidth]{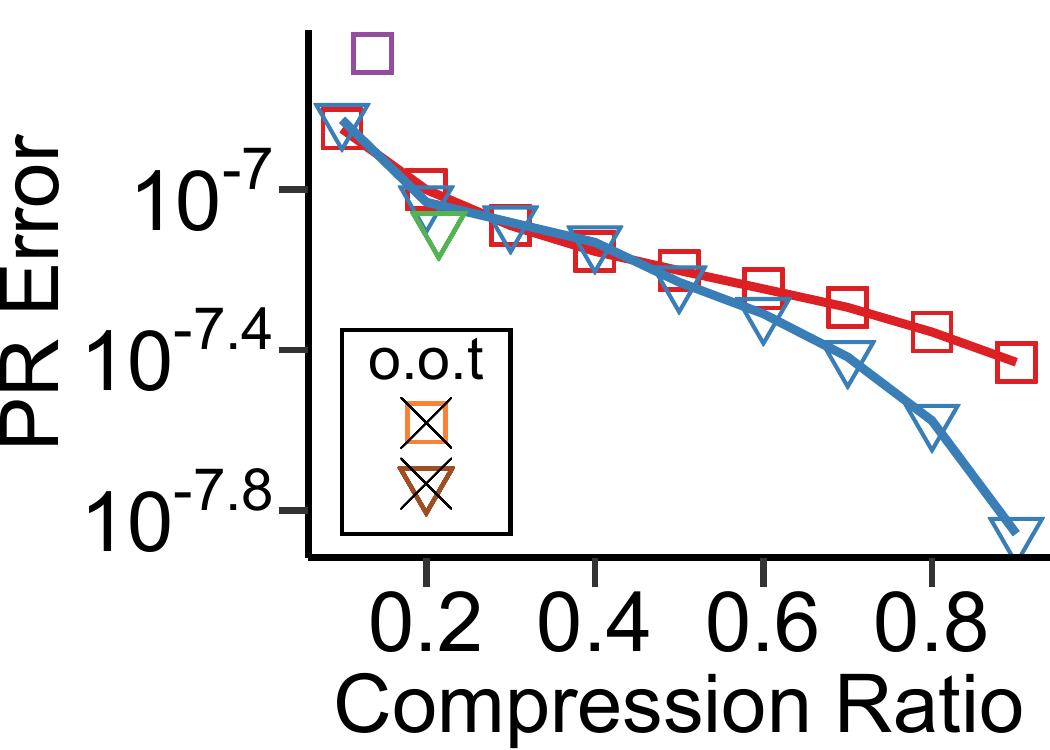}
	}
	\subfigure[DBPedia]{
		\label{fig:PageRank L1 DP}
		\includegraphics[width=0.22\textwidth]{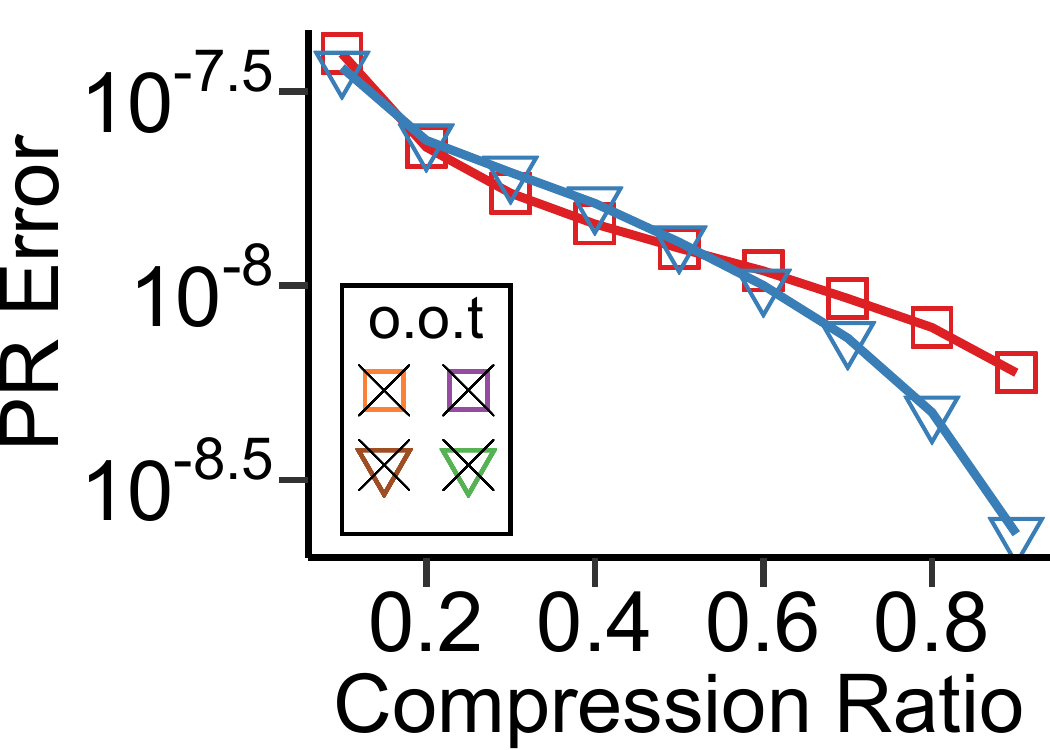}
	}  
	\subfigure[WebLarge]{
		\label{fig:PageRank L1 UK}
		\includegraphics[width=0.22\textwidth]{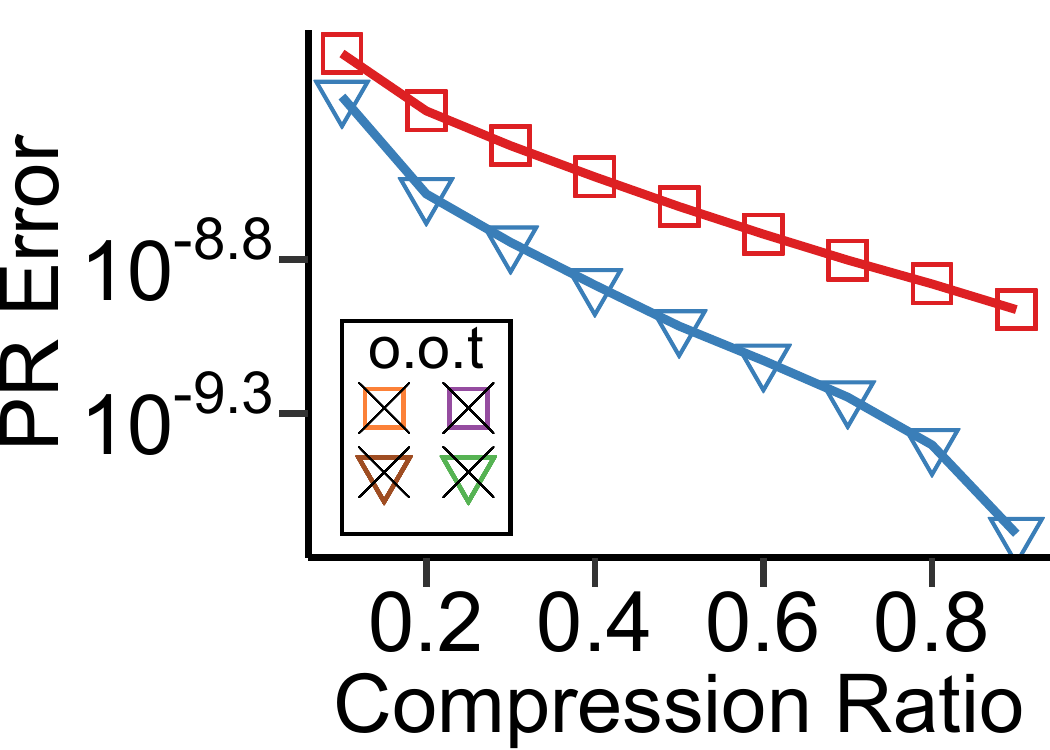}
	}  
	\\
	\textbf{Error in Node Proximity (Spec., Random Walk with Restart Scores \cite{tong2008random}):} \hfill \ \ \ \\
	\subfigure[Email-Enron]{
		\label{fig:RWR L1 EE}
		\includegraphics[width=0.22\textwidth]{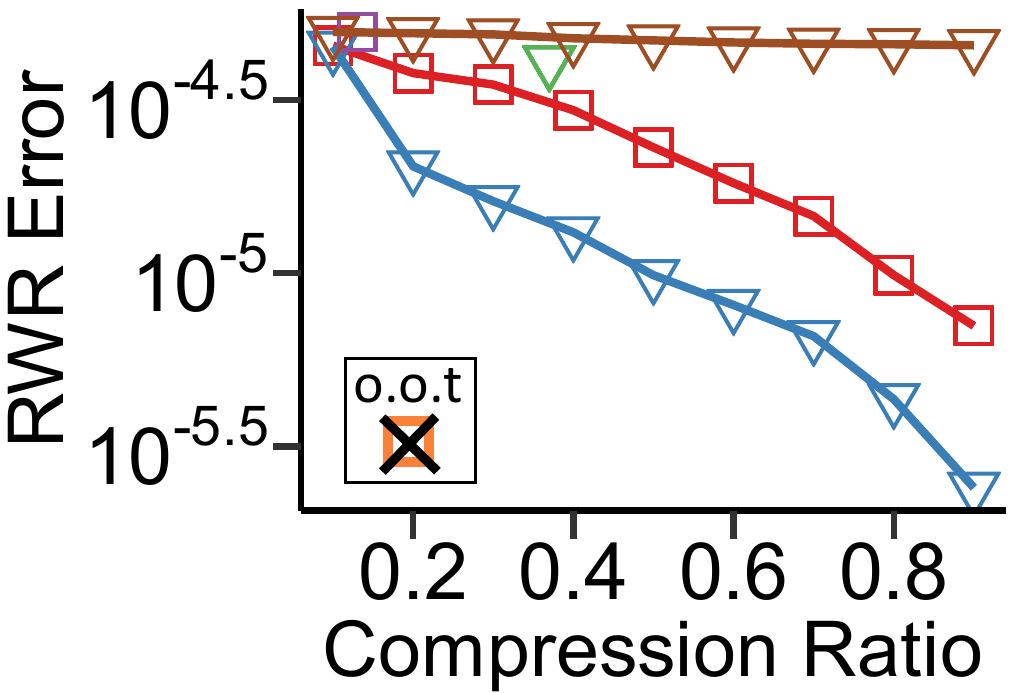}
	}
	\subfigure[DBLP]{
		\label{fig:RWR L1 DB}
		\includegraphics[width=0.22\textwidth]{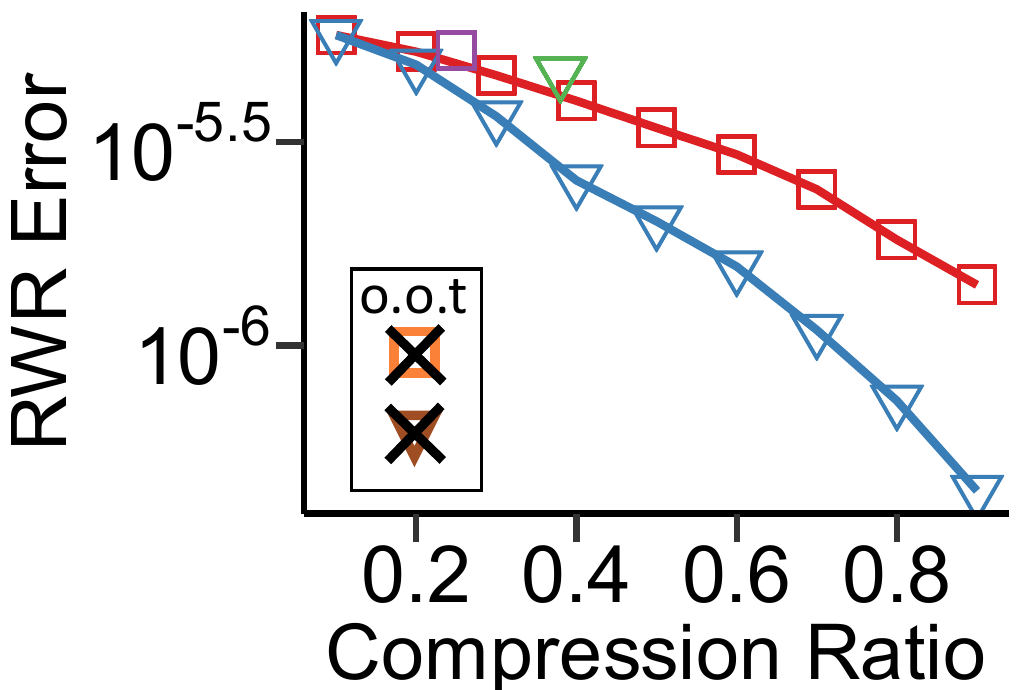}
	} 
	\subfigure[Amazon-0601]{
		\label{fig:RWR L1 A6}
		\includegraphics[width=0.22\textwidth]{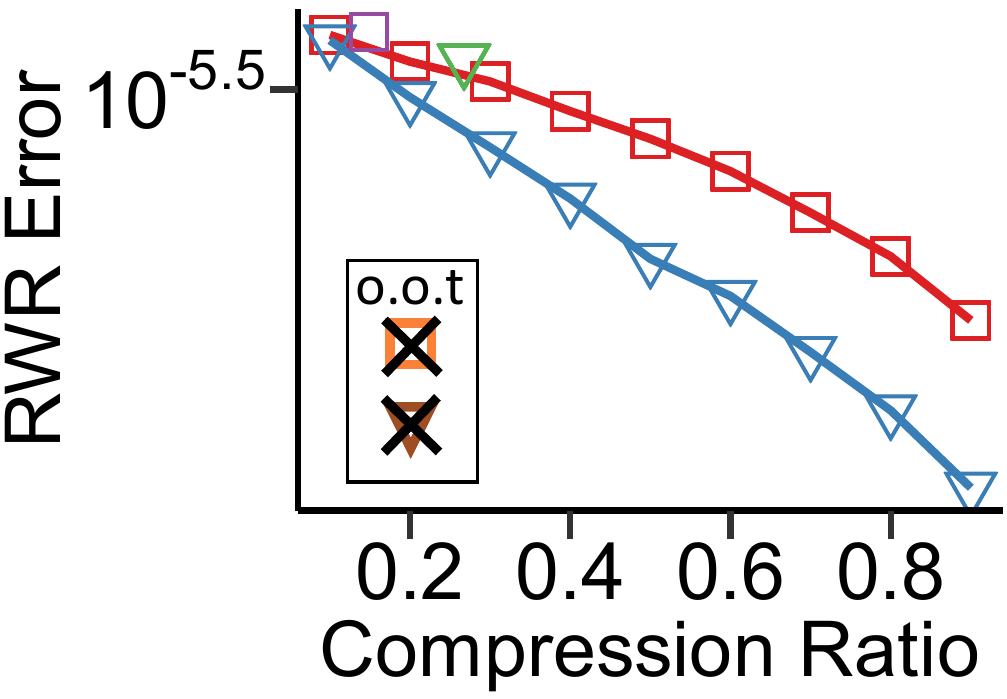}
	}
	\subfigure[WebSmall]{
		\label{fig:RWR L1 C2}
		\includegraphics[width=0.22\textwidth]{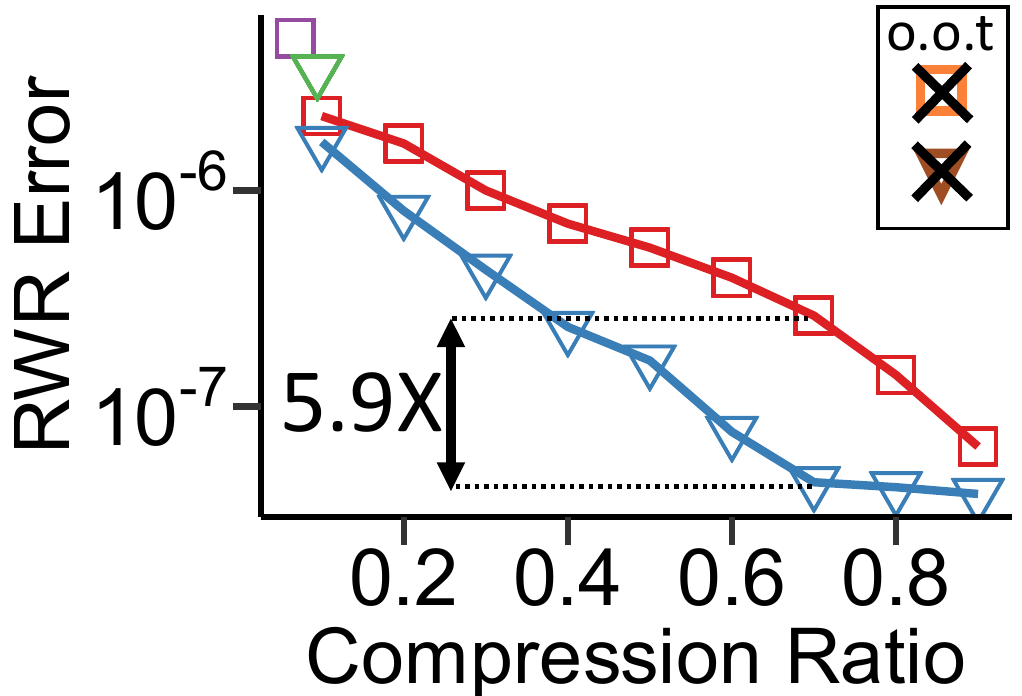}
	}
	\\
	\subfigure[Skitter]{
		\label{fig:RWR L1 SK}
		\includegraphics[width=0.22\textwidth]{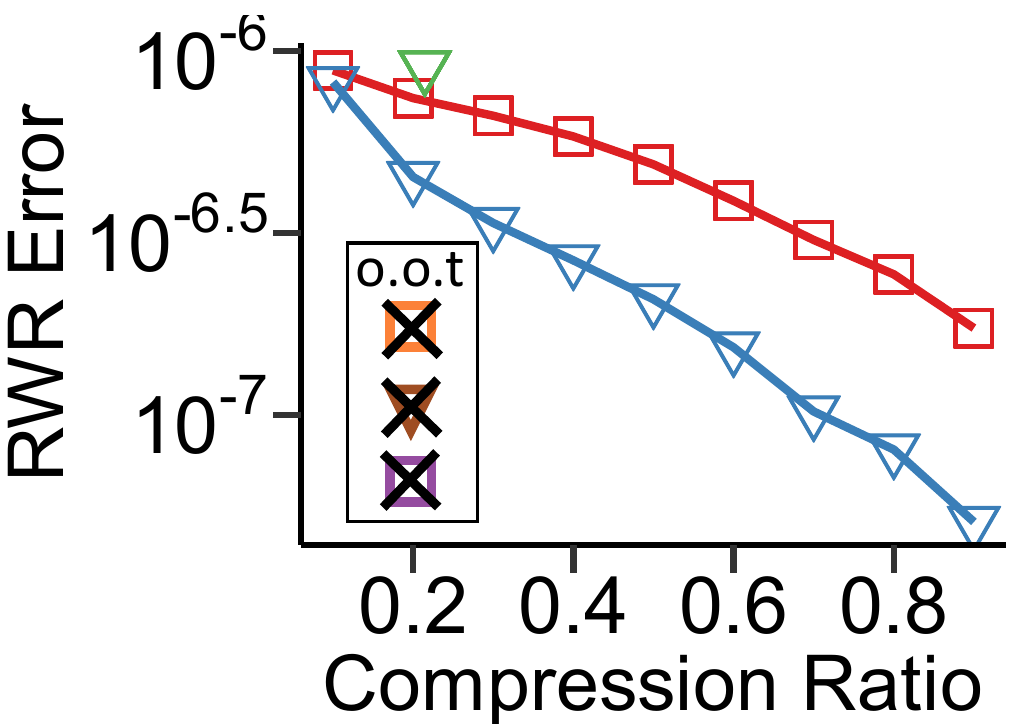}
	}
	\subfigure[LiveJournal]{
		\label{fig:RWR L1 LJ}
		\includegraphics[width=0.22\textwidth]{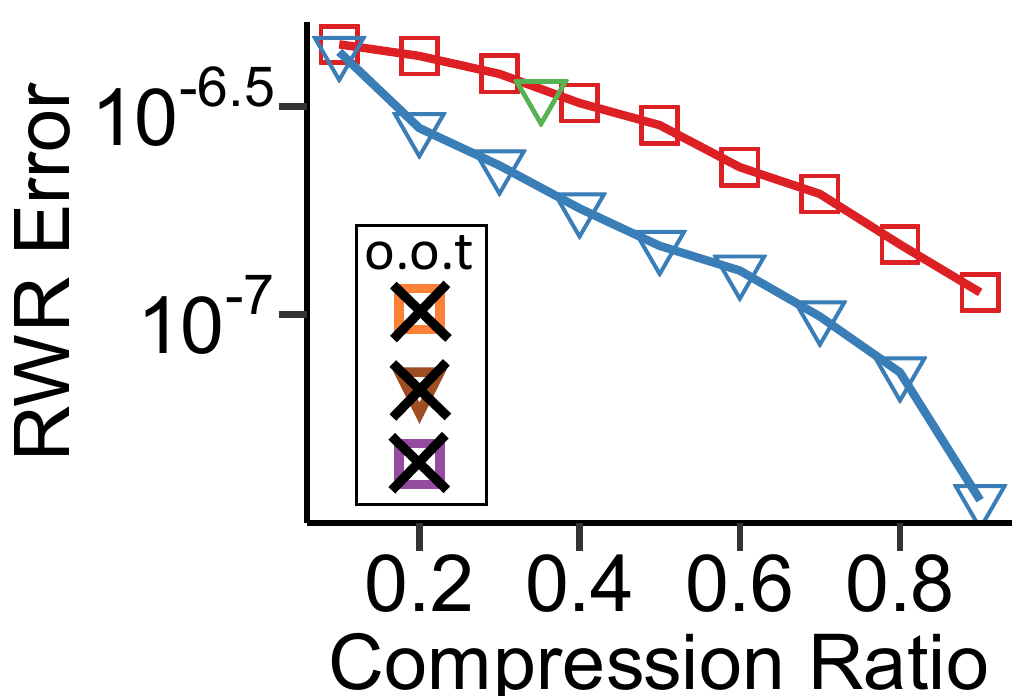}
	}
	\subfigure[DBPedia]{
		\label{fig:RWR L1 C2}
		\includegraphics[width=0.22\textwidth]{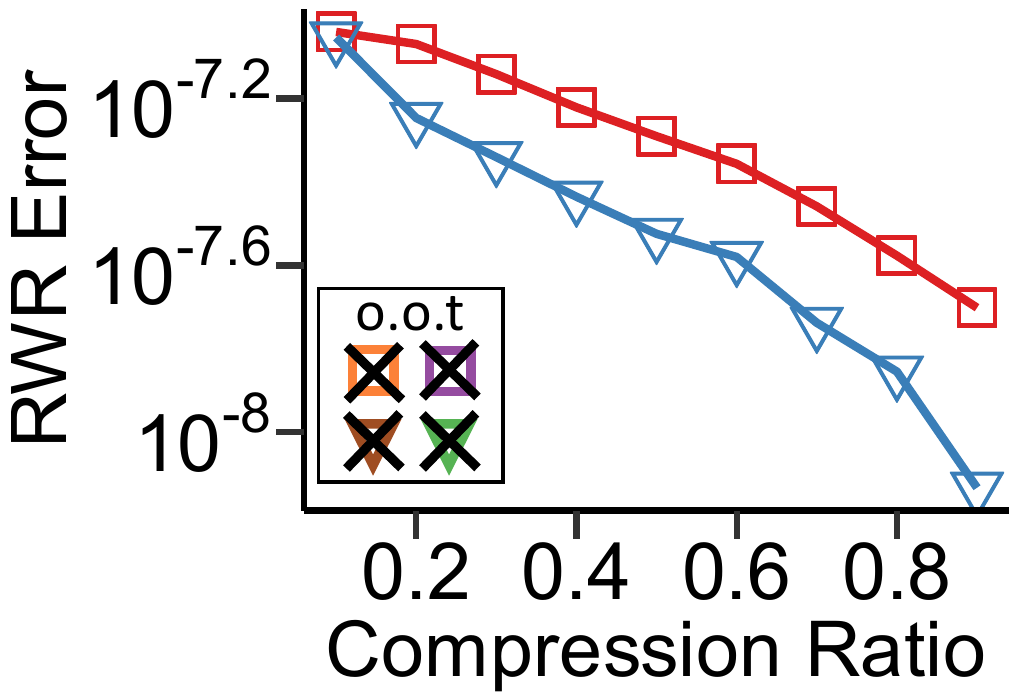}
	}  
	\subfigure[WebLarge]{
		\label{fig:RWR L1 C2}
		\includegraphics[width=0.22\textwidth]{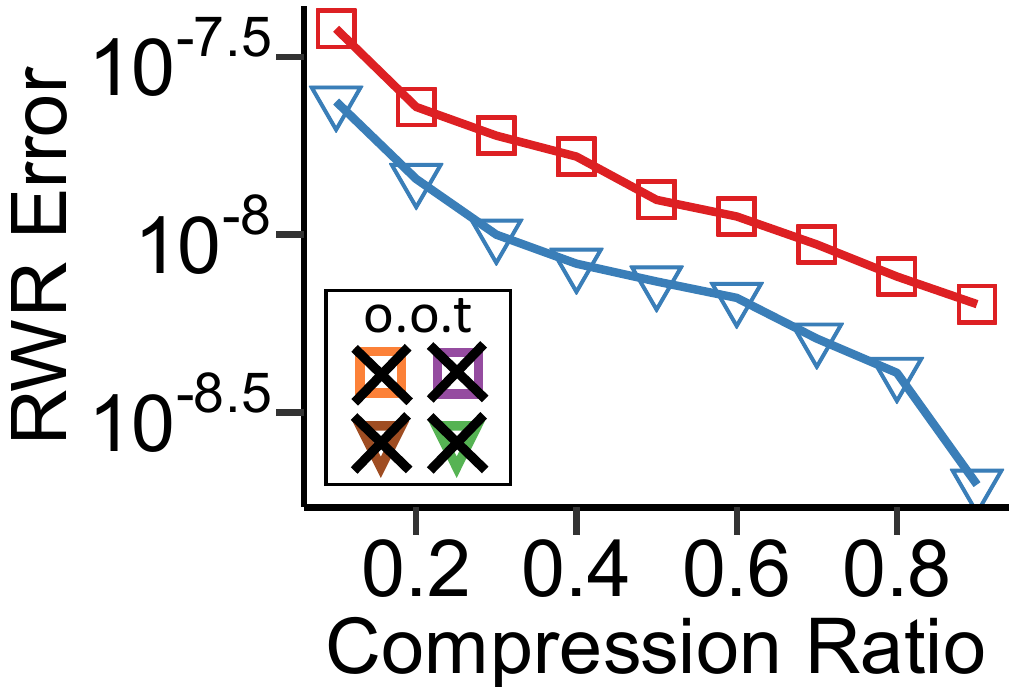}
	} 
	\\
    \caption{\label{fig:ExperimentsResult_PERWR}
    Importance of nodes and proximity between nodes are preserved more accurately in unweighted graph summarization than in weighted summarization. %in terms of Random Walk with Restart (RWR) and the number of reconstructed edges.
    \textbf{o.o.t.}: summarization or RWR computation ran out of time ($\geq$ 48 hours).
    % on summarization or query answering.
    %\textbf{o.o.r.}: out of range with too many subedges.
	}
\end{figure*}

\smallsection{Error in Node Importance:}
We used PageRank \cite{page1999pagerank} (with the damping factor $0.85$) to measure the importance of subnodes.
In Fig.~\ref{fig:ExperimentsResult_PERWR}(a)-(h), we report the sum of absolute difference between PageRank scores obtained from input and summary graphs (see Appendix~\ref{appendix:query} for how to compute PageRank scores on a summary graph).
When the compression ratios were the same, unweighted summary graphs maintained the node importance more accurately (specifically, up to $7.8\times$ when comparing \SSumM and its variant) than weighted ones.

\smallsection{Error in Node Proximity:}
We used Random Walk with Restart (RWR) \cite{tong2008random} (with the damping factor $0.95$) to measure the proximity between subnodes.
For each query node, we compute the RWR scores between the query node and the others on input and summary graphs, and we compute the sum of absolute difference (see Appendix~\ref{appendix:query} for how to compute the RWR scores on a summary graph).
In Fig.~\ref{fig:ExperimentsResult_PERWR}(i)-(p), we report the difference averaged over $100$ randomly-sampled query nodes. 
Unweighted summary graphs preserved the proximity between nodes more accurately (specifically, up to $5.9\times$ when comparing \SSumM and its variant) than weighted ones, when the compression ratios were the same.
%\red{Unweighted summary graphs preserved the proximity scores of input graphs more accurately (specifically, up to $5.9\times$ when comparing \SSumM and its variant) than weighted ones, when the compression ratios were the same.}
% 

\begin{figure*}[t!]
	\centering
	\includegraphics[width=0.7\linewidth]{./ExperimentResult/legend.pdf} \\
	\subfigure[Email-Enron]{
		\label{fig:RS L1 EE}
		\includegraphics[width=0.22\textwidth]{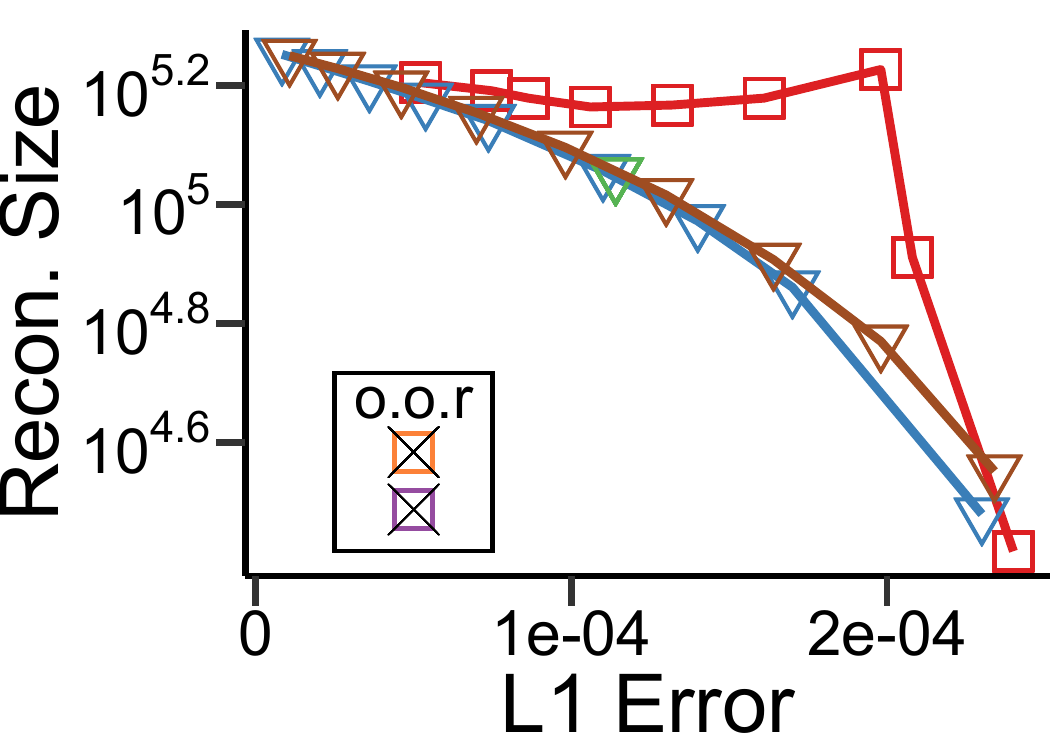}
	}
	\subfigure[DBLP]{
		\label{fig:RS L1 DB}
		\includegraphics[width=0.22\textwidth]{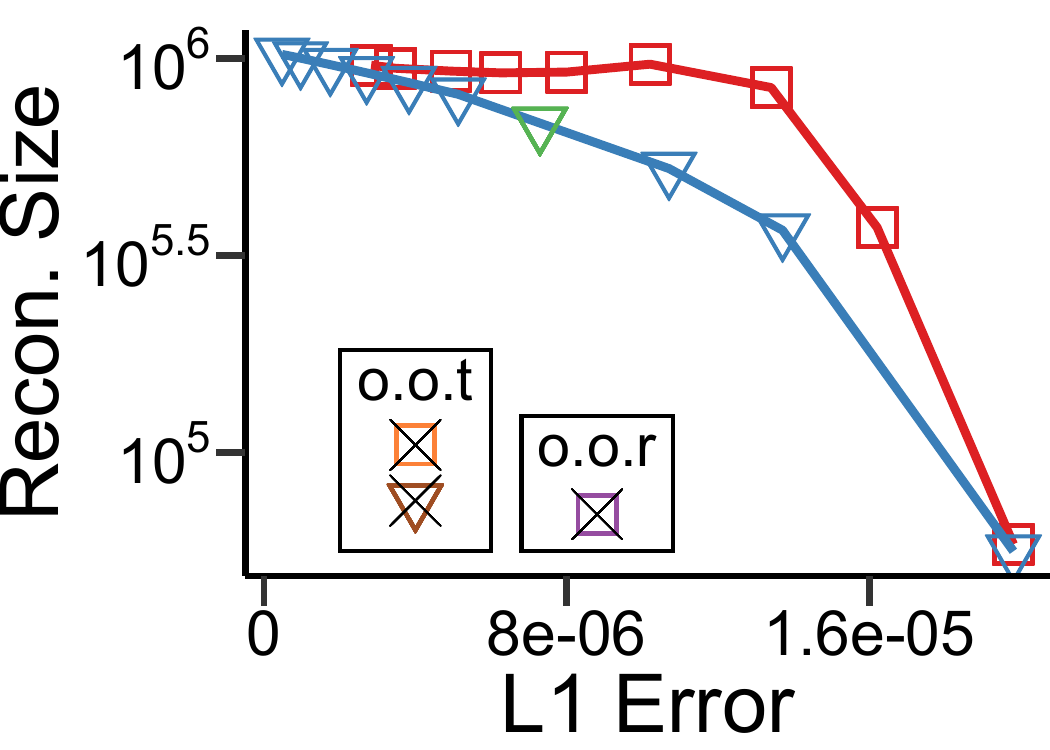}
	} 
	\subfigure[Amazon-0601]{
		\label{fig:RS L1 A6}
		\includegraphics[width=0.22\textwidth]{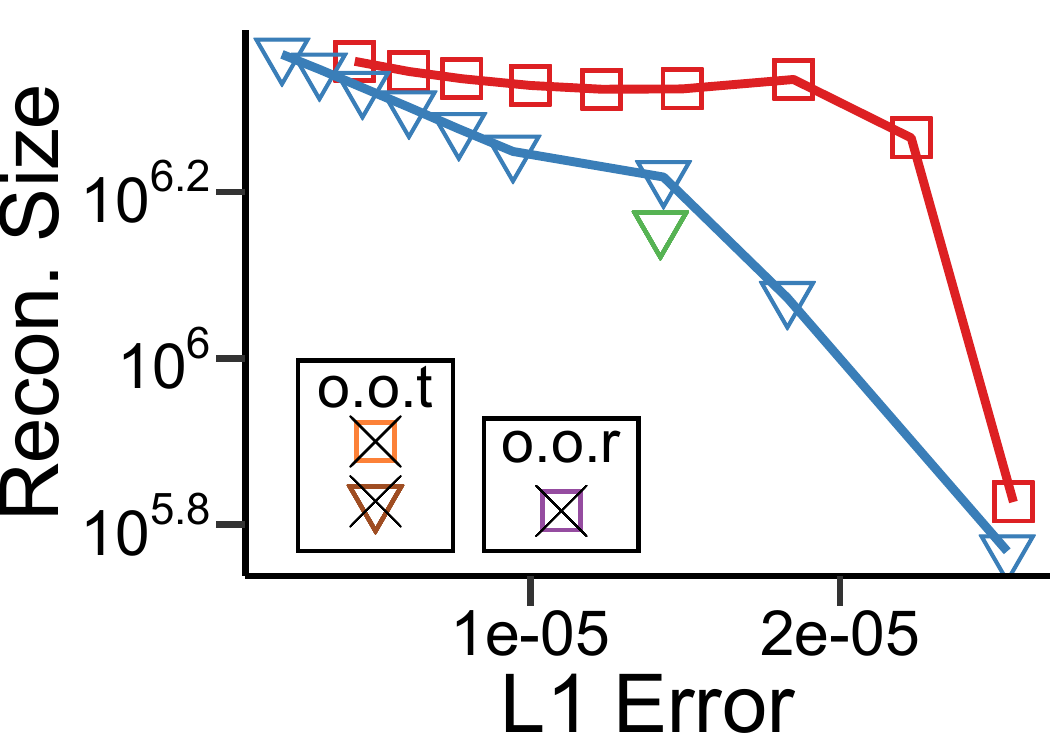}
	}
	\subfigure[WebSmall]{
		\label{fig:RS L1 C2}
		\includegraphics[width=0.22\textwidth]{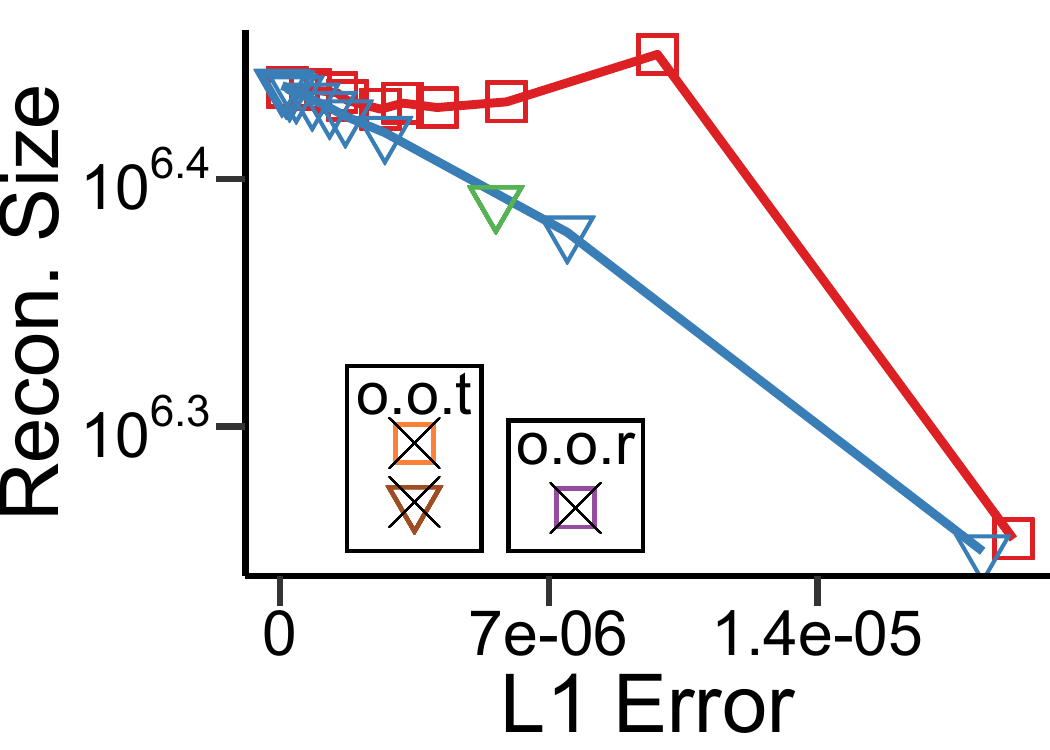}
	}
	\\
	\subfigure[Skitter]{
		\label{fig:RS L1 SK}
		\includegraphics[width=0.22\textwidth]{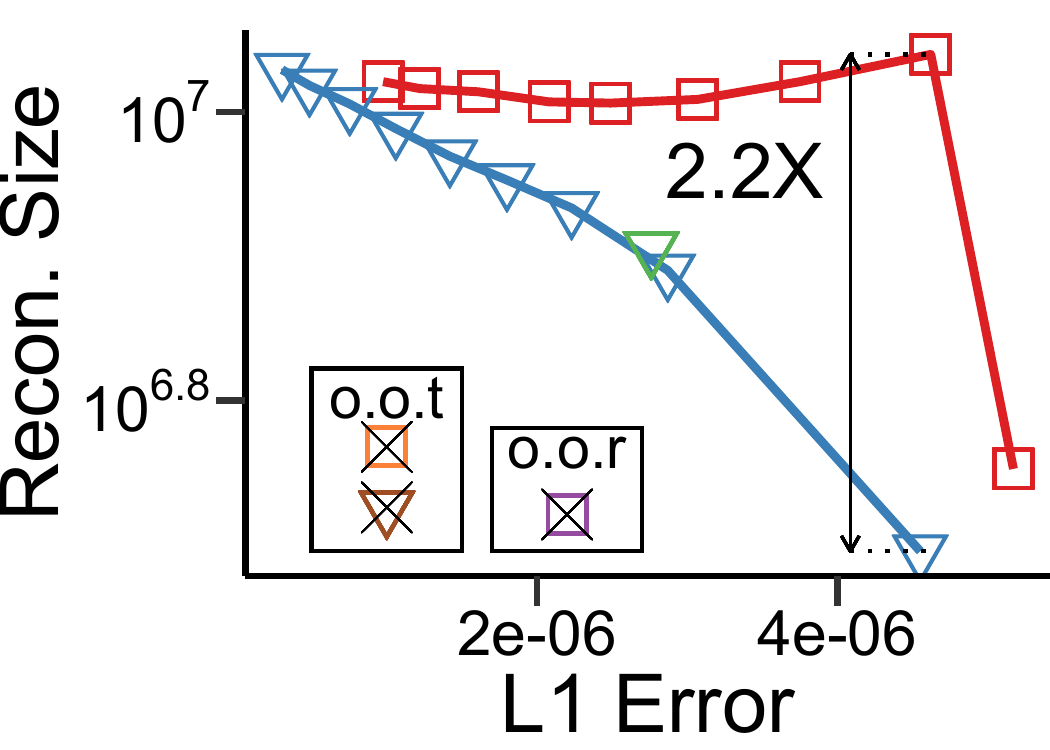}
	}
	\subfigure[LiveJournal]{
		\label{fig:RS L1 LJ}
		\includegraphics[width=0.22\textwidth]{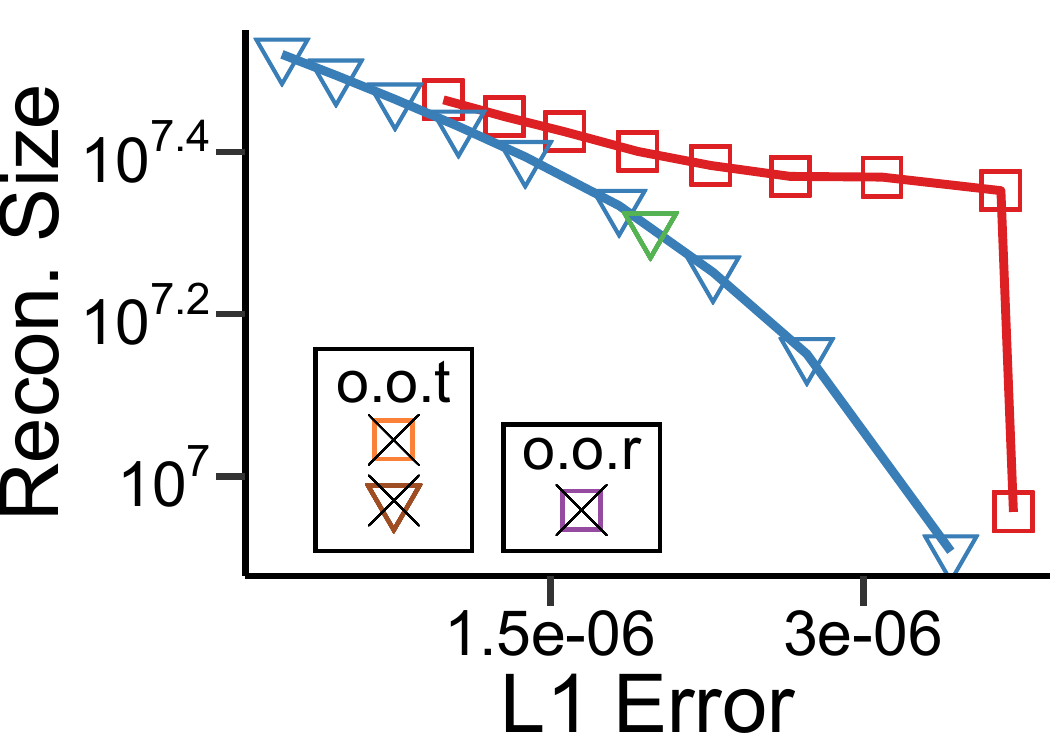}
	}
	\subfigure[DBPedia]{
		\label{fig:RS L1 C2}
		\includegraphics[width=0.22\textwidth]{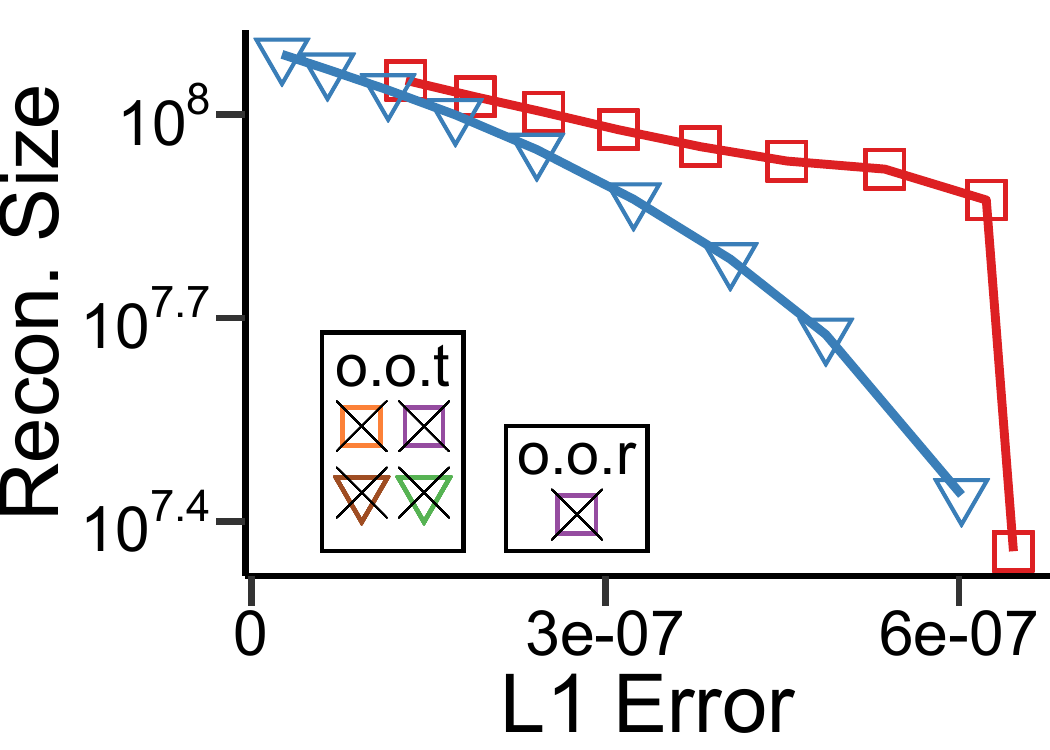}
	}  
	\subfigure[WebLarge]{
		\label{fig:RS L1 C2}
		\includegraphics[width=0.22\textwidth]{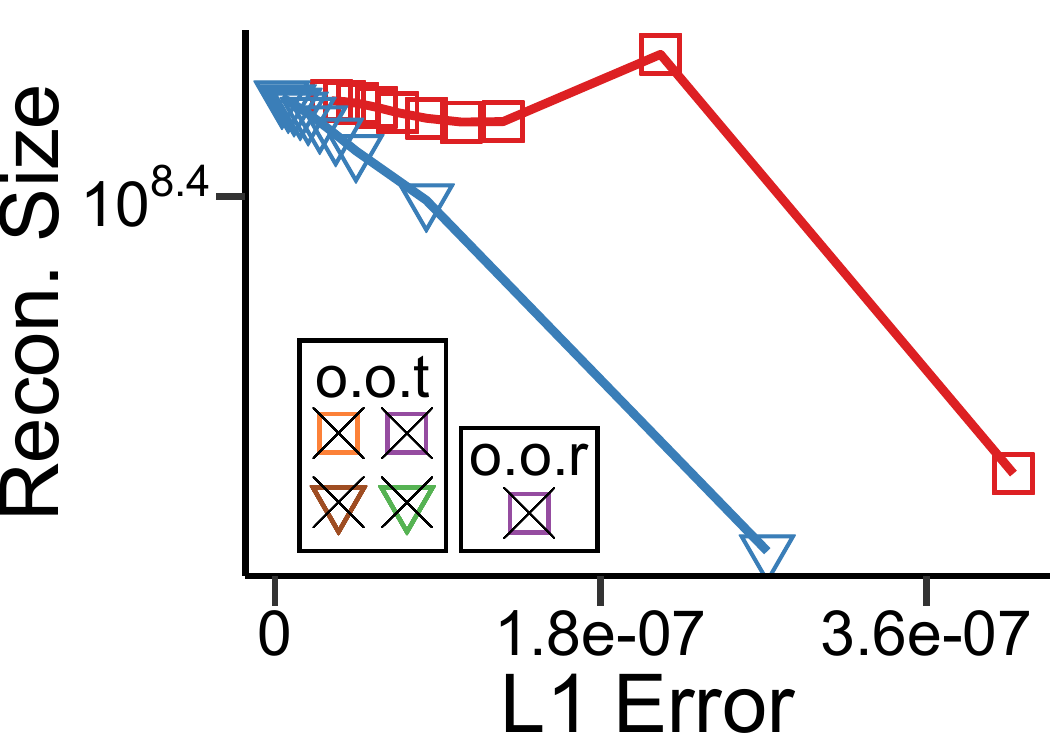}
	} 
	\caption{\label{fig:ExperimentsResult_NUMRE}
    When reconstruction errors are similar, more concise graphs are reconstructed from unweighted summary graphs than from weighted summary graphs.
    \textbf{o.o.t.}: out of time ($\geq$ 48 hours). \textbf{o.o.r.}: out of range with too many subedges.
	}
\end{figure*}

\smallsection{Size of Reconstructed Graphs:}
As shown in Fig.~\ref{fig:ExperimentsResult_NUMRE},
when $L_{1}$ reconstruction errors were similar, graphs reconstructed from unweighted summary graphs had significantly fewer (specifically, up to $2.2\times$ fewer when comparing \SSumM and its variant) subedges than those reconstructed from weighted ones.
When reconstruction errors are similar, fewer reconstructed edges, which lead to faster query processing (see \cite{riondato2017graph} and Appendix~\ref{appendix:query} for examples), are preferred.

\section{Discussion: Why Can Edge Weights be Harmful?}
\label{sec:analysis}
As answers to this question, we provide an example in Figure~\ref{fig:SummarizationandReconstruction}, and we prove in Theorem~\ref{theorem:l1} that at least when the $L_{1}$ reconstruction error is the objective, the superedge weight function $\WeightFunction$ is not useful and even harmful. The theorem, however, is not generalized to other objectives.

\begin{theorem}\label{theorem:l1}
Consider a graph $G$ and its weighted summary graph $\WeightedSummaryGraph = (S, P, \WeightFunction)$. Assume $\WeightFunction$ is not fixed but variable.
When $RE_{1}(A, \ReconstructedAdjacencyMatrix)$ is minimized, for each superedge $\{A,B\} \in P$, the weight $\frac{\WeightFunctionAB}{\PIAB}$ of subedges reconstructed from it is either $1$ or $0$, just as in Eq.~\eqref{eq:unweightedAdjValue}, where an unweighted summary graph is used.
%Note that in such a case, the reconstructed graph has either $0$ or $1$ as edge weights, as if it is reconstructed from an unweighted summary graph.
%For every superedge $P \subseteq$ $S \choose 2$, \red{encoding} them without weights minimizes L1 reconstruction error (i.e., $RE_{1}$(A, $\ReconstructedAdjacencyMatrix$)) than \red{encoding} with weights. 
\end{theorem}

\begin{proof}
%\begin{align}
%	RE_{1}(A, \ReconstructedAdjacencyMatrix) %& = \sum_{\mathclap{\{i,j\} \in V \times V}} |A_{ij} - A'_{ij}|\qquad\nonumber\\
%     = \ \ \sum_{\mathclap{\{i,j\}:\{S_i,S_j\} \in P}} \ \ |A_{ij} - A'_{ij}| + \ \  \sum_{\mathclap{\{i,j\}:\{S_i,S_j\} \notin P}} \ \ |A_{ij}| %
%	\label{eq:l1error_divide}
%\end{align}
%where $\Pi_S :=$ $S \choose 2$ $\cup \{\{X,X\}: X \in S\}$. For each superedge, L1 reconstruction error gain is as follows:
%Since the second term on the right side does not depend on $\WeightFunction$, we focus on the first term on the right side, which can be written as 
The $L_{1}$ reconstruction error can be written as follows:
\begin{equation}
	RE_{1}(A, \ReconstructedAdjacencyMatrix) \nonumber %= \sum_{\{i,j\} \in V \times V} |A_{ij} - A'_{ij}|\qquad\nonumber\\
	= \sum_{\{A,B\} \in P}\sum_{\{i,j\} \in \PIAB} |A_{ij} - A'_{ij}| + \sum_{\{A,B\} \notin P}\sum_{\{i,j\} \in \PIAB} |A_{ij}| ,
	\label{eq:l1error_divide}
\end{equation}

Since the second term on the right side does not depend on $\WeightFunction$, we focus on the first term where

% \begin{equation}
%     \sum_{\mathclap{\ \{i,j\} \in \PIAB}}  |A_{ij} - A'_{ij}| %= \sum_{\{i,j\} \in \PIAB} \left|A_{ij} - \frac{\WeightFunctionAB}{\PIAB}\right|  \nonumber \\ 
%     = \EAB\left|1- \frac{\WeightFunctionAB}{\PIAB}\right| + (\PIAB - \EAB)\left|0-\frac{\WeightFunctionAB}{\PIAB}\right|,\label{eq:l1error_edge}
% \end{equation}
\begin{equation}
    \sum_{\ \{i,j\} \in \PIAB}  |A_{ij} - A'_{ij}| %= \sum_{\{i,j\} \in \PIAB} \left|A_{ij} - \frac{\WeightFunctionAB}{\PIAB}\right|  \nonumber \\ 
    = \EAB\left|1- \frac{\WeightFunctionAB}{\PIAB}\right| + (\PIAB - \EAB)\left|0-\frac{\WeightFunctionAB}{\PIAB}\right|.\label{eq:l1error_edge}
\end{equation}
Note that Eq.~\eqref{eq:l1error_edge} is strictly larger when $\WeightFunctionAB > \PIAB$ than when $\WeightFunctionAB = \PIAB$.
Moreover, Eq.~\eqref{eq:l1error_edge} is strictly larger when $\WeightFunctionAB < 0$ than when $\WeightFunctionAB = 0$.
Thus, for the purpose of minimization, we can focus on when $\WeightFunctionAB \in [0, \PIAB]$, and thus Eq.~\eqref{eq:l1error_edge} can be rewritten as
%where $\alpha$ denotes the reconstructed adjacency matrix value, differing depending on the usage of weight. Without weight, $\alpha$ are either 0 or 1. For using weight, $\alpha$ are either 0 or $\frac{\WeightFunctionAB}{\PIAB}$. Since the domain of $\alpha$ is $0 \leq \alpha \leq 1$, we can simplify the Eq.~\eqref{eq:l1error_edge} as follows:
\begin{equation}
     \sum_{\{i,j\} \in \PIAB} |A_{ij} - A'_{ij}| = \EAB + \frac{\WeightFunctionAB}{\PIAB} (\PIAB - 2\EAB). \label{eq:l1error_final}
\end{equation}

%
%\begin{equation}
%    \sum_{\{i,j\} \in X \times Y} ||A_{ij} - A'_{ij}||_{1} := \EAB + \alpha(\PIAB - 2\EAB).\label{eq:l1error_final}
%\end{equation}
%Since Eq.~\eqref{eq:l1error_final} is a linear function, which have an $\alpha$ as a variable, we divide slope of this function $\PIAB - 2\EAB$ into 2 cases and show that our claim holds.
% More clear explanation needed... largest value of unweight is 1 while for using weight, it is M (0 <= M <= 1). Thus for cases when M is not 1, unweight superedge minimizes the L1 reconstruction gain the most.
We consider two cases depending on the sign of $\PIAB - 2\EAB$.
\begin{itemize}[leftmargin=*]
		\item {\bf Case 1.} $\PIAB < 2\EAB$:
		Since the derivative w.r.t. $\WeightFunctionAB$ is negative between $0$ and $\PIAB$, Eq.~\eqref{eq:l1error_final} is minimized when $\WeightFunctionAB=\PIAB$, i.e., when $A'_{ij}$ is $1$.
		%  is minus, $\alpha$ should be the largest value to minimize the L1 reconstruction error gain. The largest $\alpha$ value of unweight is 1 while for using weight, it is $\frac{\WeightFunctionAB}{\PIAB}$ ($0 \leq \frac{\WeightFunctionAB}{\PIAB} \leq 1$). Thus for cases when $\frac{\WeightFunctionAB}{\PIAB}$ is not 1, unweight superedge minimizes the L1 reconstruction gain the most. 
		\item {\bf Case 2.} $\PIAB \geq 2\EAB$:
		Since the derivative w.r.t. $\WeightFunctionAB$ is non-negative between $0$ and $\PIAB$, Eq.~\eqref{eq:l1error_final} is minimized when $\WeightFunctionAB=0$, i.e., when $A'_{ij}$ is $0$, 
		
		%$\PIAB \geq 2\EAB$: Since the slope is plus, $\alpha$ should be the smallest value to minimize the L1 reconstruction error gain. Thus, in this case, there should be no superedge representing the connection between supernode $X$ and $Y$, meaning the value $\alpha$ is 0. %which does not differ using weights or not.
\end{itemize}

Therefore, when Eq.~\eqref{eq:l1error_edge} is minimized, for each superedge $\{A,B\} \in P$, the weight of subedges reconstructed from it (i.e., $\frac{\WeightFunctionAB}{\PIAB}$) is either $1$ or $0$, as in the unweighted model.  \hfill $\blacksquare$
\end{proof}

\section{Conclusion and Future Directions}\label{sec:conclusions}

In this work, we conducted a systematic comparison between two extensively-studied graph summarization models with and without superedge weights. To this end, we extended three search algorithms to both models (Algorithms~\ref{alg:static}-\ref{alg:dynamic} and Table~\ref{tab:algorithmDetail}) and compared their outputs from eight real-world graphs in five aspects (Figs.~\ref{fig:ExperimentsResult_L1L2}-\ref{fig:ExperimentsResult_NUMRE}).
Our empirical comparison revealed a surprising finding that removing superedge weights leads to significant improvements in all five aspects, as in the example in Fig.~\ref{fig:SummarizationandReconstruction}.
Then, we developed a theoretical analysis to shed light on this counterintuitive observation (Theorem~\ref{theorem:l1}).
Noteworthy, we showed in Figs.~\ref{fig:ExperimentsResult_L1L2}-\ref{fig:ExperimentsResult_NUMRE} that \SSumM \cite{lee2020ssumm}, a state-of-the-art graph-summarization algorithm, can be improved substantially (specifically, up to $8.2\times$, $7.8\times$, and $5.9\times$ in terms of reconstruction error, error in node importance, and error in node proximity, respectively, when the compression ratio was fixed; and $2.2\times$ in terms of the size of reconstructed graphs, when the reconstruction error was similar) based on the observation.
As future work, we would like to explore (a) better superedge weighting schemes and (b) combinations of weighted and unweighted superedges.  

%delete
\smallsection{Reproducibility:} The source code and the datasets are available at \cite{appendix}.

\smallsection{Acknowledgements:}
This work was supproted by National Research Foundation of Korea (NRF) grand funded by the Korea government (MSIT) (No. NRF-2020R1C1C1008296) and Institute of Information \& Communications Technology Planning \& Evaluation (IITP) grant funded by the Korea government (MSIT) (No.2019-0-00075, Artificial Intelligence Graduate School Program (KAIST)).

%\smallsection{Reproducibility:}
%The source code and datasets used in this paper are available at \cite{appendix}.

\bibliographystyle{plain}
\bibliography{098reference.bib}

\appendix
\section{Appendix: Graph Algorithms on Summary Graphs}
\label{appendix:query}

Given a summary graph $G'$ (i.e., $\WeightedSummaryGraph$ or $\UnweightedSummaryGraph$) and a query node $u\in V$, an approximate set of neighbors of $u$ can be retrieved from $G'$ without reconstructing the entire graph, as described in Algorithm~\ref{alg:NeighborQuery}.
In other words, neighborhood queries can be answered approximately from $G'$.
A wide range of graph algorithms (e.g., DFS, BFS, PageRank, and Dijkstra's) access the input graph only through neighborhood queries, and thus they can be executed approximately on summary graphs without restoring the entire graph. See Algorithm~\ref{alg:RWR} for examples. % where the \underline{\smash{underlined lines}} are required for Random Walk with Restart but not for PageRank.

\begin{algorithm}[H]
    \small 
    \caption{\small \textbf{getNeighbors($\SummaryGraph$,$u$)}}
    \label{alg:NeighborQuery}
\DontPrintSemicolon
    \KwInput{(1) summary graph $G'$ ($\WeightedSummaryGraph$ or $\UnweightedSummaryGraph$) and (2) query subnode $u$
    }
    \KwOutput{approximate neighborhood $\hat{N}_{u}$ of $u$ with subedge weights }
    \SetAlgoLined
    $\hat{N}_{u} \leftarrow \emptyset$\\ 
    \For{$\mathbf{each}\ A$ \normalfont{where} $\{A,S_u\}\in P$}{
     %       \Comment*[r]{$S_{u}$: the supernode that contains $u$}
        \For{$\mathbf{each}\ v \in A$}{ 
            \If{$v \neq u$}{
                % add $v$ to $\hat{N}_{u}$ with weight $\frac{\WeightFunction_{AS_{u}}}{\Pi_{AS_{u}}}$ (or 1) %\blue{\frac{\WeightFunction_{AS_{u}}}{}} 
                \If{$G' = \WeightedSummaryGraph$}{
                add $v$ to $\hat{N}_{u}$ with weight $\frac{\WeightFunction_{AS_{u}}}{\Pi_{AS_{u}}}$}
               \If{$G' = \UnweightedSummaryGraph$}{
                add $v$ to $\hat{N}_{u}$ with weight $1$}  
            }
        }
        % $\hat{N}_{u} \leftarrow \hat{N}_{u} \cup U$
    }
    % $\hat{N}_{u} \leftarrow \hat{N}_{u} \setminus \{u\}$\\
    \textbf{return} $\hat{N}_{u}$ 
    % \caption{getNeighbors($\SummaryGraph$, $u$) on \blue{$\WeightedSummaryGraph$} or \red{$\UnweightedSummaryGraph$}}
\end{algorithm}

% \small 
% \red{* When the input summary graph is $\WeightedSummaryGraph$,} \\\red{$w$ is $\frac{\WeightFunction_{AS_{u}}}{\Pi_{AS_{u}}}$. Otherwise, $w$ is $1$.}\\
%  \\ 
% \red{*** one-hot(u) is the one-hot vector of} \\ \red{size $|V|$ whose $u$-th entry is $1$.}\\

\begin{algorithm}[H]
    \small
\DontPrintSemicolon
    \KwInput{(1) summary graph $\SummaryGraph$, (2) damping factor $d$, and \underline{\smash{(3) (only for RWR) query subnode $u$}}\\
    }
    \KwOutput{score vector $r^{new}\in \mathbb{R}^{|V|}$}
    $V\leftarrow \bigcup_{A\in S}A$ \\
    $r^{old}\leftarrow \textbf{0}$ ; 
    $r^{new} \leftarrow \frac{1}{|V|}\cdot \textbf{1}$  \hfill \Comment{\textbf{0} is the zero vector of size $|V|$}
    % \For{$\mathbf{each}\ B \in S$}{
    %     \For{$\mathbf{each}\ b \in B$}{
    %         $V\leftarrow V \cup \{b\}$
    %     }
    % }
    
 %   \For{$\mathbf{each}\ M \in S$}{
        % \For{$\mathbf{each}\ b \in B$}{
        %     $V\leftarrow V \cup \{b\}$
        % }
  %      $V \leftarrow V \cup M$
        % $V\leftarrow V \cup S(M)$ \Comment*[r]{$S(M)$ = set of nodes in supernode $M$}
  % }
    $q \leftarrow$ $\frac{1}{|V|}\cdot\textbf{1}$ \hfill \Comment{\textbf{1} is the one vector of size $|V|$}
    \underline{\smash{(only for RWR) $q \leftarrow \textbf{0}$; $q_u \leftarrow 1$}} \\
    %   \hfill \Comment{one-hot(u) is the one-hot vector of size $|V|$ whose $u$-th entry is $1$}
    
    % \Comment*[r]{\textbf{1} = one vector of size |V|} 
    % \blue{\textbf{q} $\leftarrow \frac{1}{|V|}\cdot\textbf{1}$}\\
    % \red{\textbf{q} $\leftarrow$ \textbf{one-hot($u$)}} \Comment*[r]{\blue{one-hot vector of size |V| that entry of the node $u$ is set to 1}}
%    \red{\textbf{q} $\leftarrow$ \textbf{one-hot($u$)}} \Comment*[r]{one-hot vector of size |V| that \textbf{q}$_{u}$ = 1}
    % \red{\textbf{q} $\leftarrow$ \textbf{one-hot$^{u=1}$}} \\
    % \Comment*[r]{\blue{one-hot vector of size |V| that entry corresponds to the node $u$ is set to 1}}
    \While{$r^{new}\neq r^{old}$}{
        $r^{old} \leftarrow r^{new}$; $r^{new} \leftarrow \textbf{0}$ \\
        \For{$\mathbf{each}\ v \in V$}{
            $\hat{N}_{v} \leftarrow$ \textbf{getNeighbors($\SummaryGraph$, $v$)} \\
            $w_{sum} \leftarrow$ sum of weights in $\hat{N}_{v}$
            \\
            \For{$\mathbf{each}$ \normalfont{neighbor} $l$ \normalfont{with weight} $w$ \normalfont{in} $\hat{N}_{v}$}{ 
              $r_{l}^{new} \leftarrow r_{l}^{new} + \frac{w}{w_{sum}}r_{v}^{old}$ 
            } 
        }
        $r^{new} \leftarrow d\cdot r^{new} + (1-d\cdot \sum_{v\in V} r_{v}^{new})\cdot\textbf{q}$ \nonumber \\
    }
    \textbf{return} $r^{new}$ 
    \caption{\small PageRank \cite{page1999pagerank} and  \underline{\smash{Random Walk with Restart (RWR)}} \cite{tong2008random} on $\SummaryGraph$}
    \label{alg:RWR}
\end{algorithm}

\end{document}

\title{Are Edge Weights in Summary Graphs Useful? - A Comparative Study (Online Appendix)} % of Graph Summarization Models}

\author{}
\institute{}{}

\maketitle              % typeset the header of the contribution

\section{Appendix: Error in Node Importance}
% \section{Appendix: Error in PageRank Scores}
% \appendix

% \subsection{Error in PageRank Scores}

\setcounter{figure}{4}

\begin{figure*}
    \centering
	\subfigure[Compression Ratio]{
		\label{fig:Appendix SN-CR}
		\includegraphics[width=0.32\textwidth]{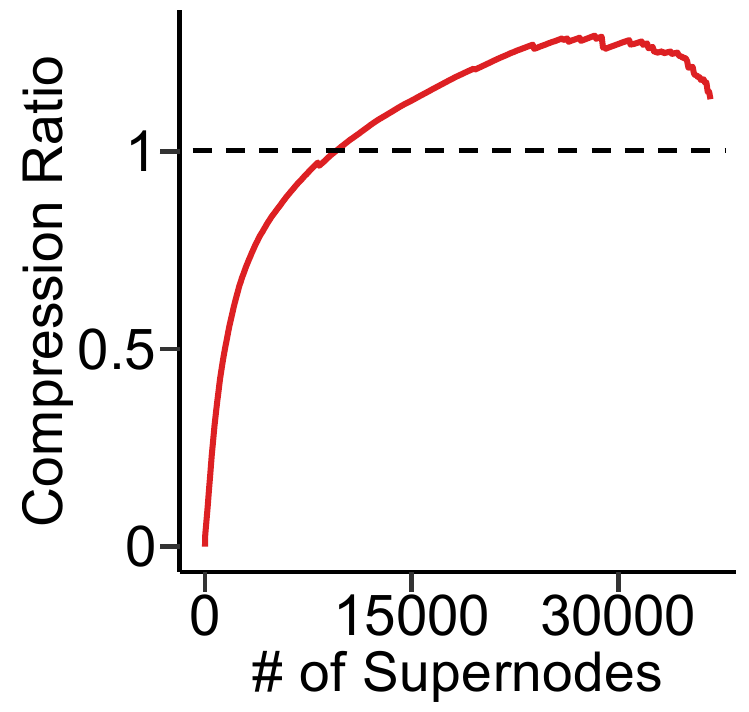}
	}
	\subfigure[PageRank Error]{
		\label{fig:Appendix SN-PRE}
		\includegraphics[width=0.32\textwidth]{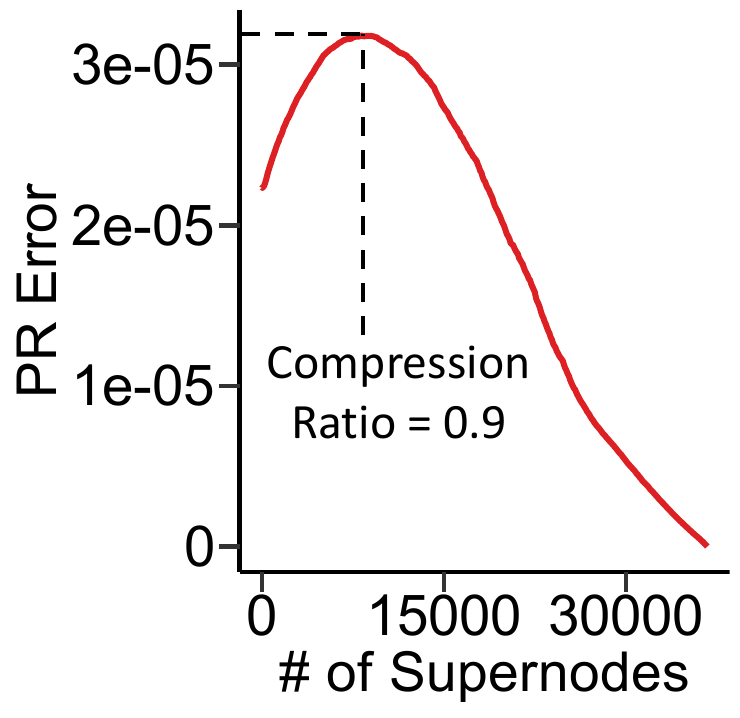}
	}
	\caption{The compression ratio and the error in PageRank scores in summary graphs obtained by \kGrass.
	\label{fig:AppendixResults}
	}
\end{figure*}

In Fig.~3(a) of the main paper, the error in PageRank scores decreases as we compress the input graph more using \kGrass.
For further analysis of this counterintuitive phenomenon, we measure the compression ratio and the error in PageRank scores as we change the number of supernodes in the summary graph obtained by \kGrass.

As seen in Fig.~\ref{fig:Appendix SN-CR}, the compression ratio is greater than $1$ (i.e., the summary graph is bigger than the input graph in terms of the number of bits) until the number of supernodes reaches about $10,000$.
This is because, in the early stage of \kGrass, the number of bits required for storing superedge weights increases faster than the number of bits saved by merging nodes.
After merging supernodes sufficiently many times, the compression ratio begins to decrease.
As seen in Fig.~\ref{fig:Appendix SN-PRE}, the error in PageRank scores first increases as we decrease the number of supernodes, but the error starts to decrease after merging supernodes sufficiently many times.
Fig.~3(a) of the main paper shows the results for a range of the number of supernodes where both the compression ratio and the error in PageRank scores decrease as we merge more supernodes.